\documentclass[11pt]{article}
\usepackage{makeidx}
\usepackage[dvipsnames,svgnames,table]{xcolor}
\usepackage{epstopdf}
\usepackage{epsfig}
\usepackage{textcomp}
\usepackage{hyperref}
\usepackage{amsmath}
\usepackage{amssymb}
\usepackage{multirow}
\usepackage{multicol}
\usepackage{booktabs}
\usepackage{lastpage}
\usepackage{hyperref} 
\usepackage{graphicx}
\usepackage{enumerate}
\usepackage{threeparttable}
\usepackage{float}
\usepackage{array}
\usepackage{cite}
\newcommand{\sgn}{\text{sgn}}
\begin{document}
\setcounter{page}{1}
\begin{center}
\large{\textbf{Enabling microstructural changes of FCC/BCC alloys in 2D dislocation dynamics}}
\end{center}
\bigskip
\begin{center}
\noindent\textbf{Ahmet Ilker Topuz}\\
\medskip
Materials innovation institute (M2i), 2600 GA Delft, The Netherlands\\
E-mail address: aitopuz@gmail.com, i.topuz@m2i.nl\\
Tel: +31 (0)15 278 2535\\
Fax: +31 (0)15 278 2591\\
\begin{abstract}
Dimension reduction procedure is the recipe to represent defects in two dimensional dislocation dynamics according to the changes in the geometrical properties of the defects triggered by different conditions such as radiation, high temperature, or pressure. In the present study, this procedure is extended to incorporate further features related to the presence of defects with a special focus on face-centered cubic/body-centered cubic alloys used for diverse engineering purposes. In order to reflect the microstructural state of the alloy on the computational cell of two dimensional dislocation dynamics, the distribution of the multi-type defects over slip lines is implemented by using corresponding strength and line spacing for each type of defect. Additionally, a simple recursive incremental relation is set to count the loop accumulation on the precipitates. In the case of continuous resistance against the motion of edge dislocations on the slip lines, an expression of friction is introduced to see its contribution on the yield strength. Each new property is applied independently on a different material by using experimental information about defect properties and grain sizes under the condition of plain strain deformation: both constant and dynamically increasing obstacle strength for precipitate coarsening in prime-aged and heat-treated copper-chromium-zirconium, internal friction in tantalum-2.5tungsten, and mixed hardening due to the presence of precipitates and prismatic loops in irradiated oxide dispersion strengthened EUROFER with 0.3$\%$ yttria.
\end{abstract}
\end{center}
\bigskip

\section{Introduction}
Combination of one or more metals or non-metals is an age-old technique to improve the mechanical properties such as yield strength. In addition to the alloying, the tensile response of the materials is optimized by the use of supplementary processes like annealing or quenching which result in the evolution of the microstructure. In extreme environments, the initial state of the materials may not be maintained due the operating conditions, and the performance of the materials may vary according to the perturbation in the intrinsic properties. At this point, the design of materials for fusion or generation IV reactors becomes an industrial necessity in order to deal with technological problems. 

Experimental investigations on various nuclear alloys including but not limited to copper alloys~\cite{Woolhouse1971,Satoh1988,Singh1996,CuCrZr}, austenitic steels~\cite{Holmes1968, Lucas1993, Zinkle1993, Hashimoto1999,Jiao2010}, and ferritic/martensitic steels~\cite{Schaeublin2002,Klueh2007,Tanigawa2009,Tanigawa2011} have been widely performed in order to understand the relation between the evolution of defects and the constitutive behavior. As reported by these experiments, precipitates are commonly detected in FCC/BCC alloys~\cite{Woolhouse1971, Singh1996,CuCrZr, Lucas1993, Hashimoto1999,Klueh2007,Tanigawa2009} at different irradiation levels~\cite{Singh1996,Hashimoto1999,Tanigawa2009} and periods of heat treatment~\cite{CuCrZr}, and significant changes due to these processes in the geometrical properties of precipitates result in the variation of mechanical properties. There exists also certain types of defects which are induced by irradiation in specific crystal structures. While the formation of stacking-fault tetrahedra (SFT) occurs in some face-centered cubic (FCC) materials~\cite{Satoh1988,Singh1996}, dislocation loops are the major irradiation-induced defects in body-centered cubic (BCC) materials~\cite{Tanigawa2009,Cockeram2011}. At elevated temperatures (e.g. $>$573 K~\cite{Cockeram2011} or $>$623 K~\cite{Lucas1993}), the population of voids is significant and contributes to the change of the tensile properties. 
 
When dislocation-barrier interaction is considered, precipitates and related mechanisms in the alloys have been studied in 3D dislocation dynamics (DD)~\cite{Monnet2006, Bako2007, Queyreau2010, Robertson2011}. The themes of some exemplary works dedicated for the simulations of precipitates in 3D DD are precipitation-induced strengthening in a Zr-$1\%$ Nb alloy proposing a mixture law~\cite{Monnet2006}, interactions between dislocations and  $\rm Y_{2}O_{3}$ particles in PM2000 single crystals~\cite{Bako2007}, combination of Orowan mechanism and forest hardening  in reactor pressure vessel (RPV)  steels~\cite{Queyreau2010}, and utilization of  impenetrable facets for incoherent oxide particles together with shear-able facets for irradiation induced sessile dislocation loops in oxide dispersion strengthened (ODS) materials~\cite{Robertson2011}. Although 3D DD delivers an illuminating description about the motion of dislocations, it is considered as an expensive method of simulation in terms of computation time. Moreover, current limitations such as simulation volumes far below the grain size of many engineering materials imply that 3D DD may be desirable for the cases where 3D setup is indispensable.

In the present study, additional features based on dimension reduction procedure (DRP)~\cite{Topuz} in 2D DD framework~\cite{main} are introduced in order to reflect the changes in the microstructure of FCC/BCC alloys. In section~\ref{Dimension reduction of defects}, a brief description for the content of dimension reduction is given. Then, in section~\ref{Additional features for alloys}, representation of multi-type defects, internal friction, and dynamically increasing obstacle strength is shown. Section~\ref{BCC slip systems and their effective Burgers vector magnitude} describes BCC slip systems according to the suggested configuration stated in another study~\cite{Ricebcc} for plain strain deformation and explains the calculation of effective Burgers vector magnitude~\cite{Wang2009} in the corresponding slip systems. Finally, in section~\ref{Results and discussion}, each new property is applied separately on a different alloy by using experimental information about the geometrical properties of defects under the condition of plain strain deformation: both constant and dynamically increasing obstacle strength for precipitate coarsening in prime-aged and heat-treated FCC CuCrZr~\cite{CuCrZr}, internal friction in BCC Ta-2.5W~\cite{Khan1999}, and mixed hardening in irradiated BCC ODS EUROFER with 0.3$\%$ yttria~\cite{Ramar2007, Heintze2011}.
\section{Dimension reduction of defects}
\begin{figure}[H]
\centering
\includegraphics[width=16cm]{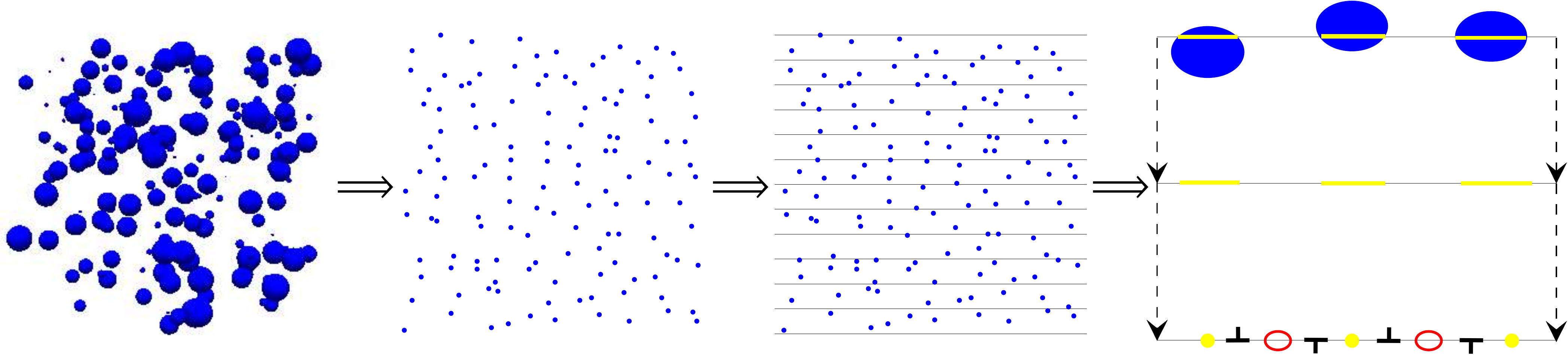}
\caption{Transition of 3D geometrical properties of defects into 1D information for slip lines on which dislocations \textbf{$(\bot)$} are generated from Frank-Read (FR) sources \textbf{$(\bigcirc)$}, can glide and be pinned/depinned at obstacles \textbf{$(\bullet)$}.}
\label{2Dscheme}
\end{figure}

DRP~\cite{Topuz} of defect properties in 2D DD~\cite{main} is the transition of 3D geometrical information to the basis of slip lines by using the principles of fractional conservation~\cite{Dele,Rosi} and barrier hardening formulations such as Dispersed Barrier Hardening (DBH)~\cite{DBH} model or Bacon-Kocks-Scattergood (BKS) model~\cite{BKS}. This procedure is performed by the application of two consecutive slicing operations on the objects of a certain number, size and geometry as described in Fig.~\ref{2Dscheme}. First slicing yields an average 2D profile in terms of density and size and it permits the determination of defect strength assuming that 2D average spacing is preserved inside the geometrical cell. At the end of the latter slicing, all 3D objects are reduced to the segments as a result of intersections between 2D objects and slip lines. In this manner, line density which is defined as the number of segments per slip line is obtained to represent defects in 2D DD. For practical reasons, in lieu of segments, points are distributed over all slip lines. Moreover, prismatic dislocation loops are considered as representative segments on the lateral cross section of the material, and a special treatment that counts the number of interactions between a mesh of both parallel and perpendicular lines and segments of prismatic loops is applied~\cite{Topuz}.  
\label{Dimension reduction of defects}
For each type of defect, DRP yields an ordered pair of line density and strength varying with 3D size and density at every level of any process. Density and spacing formulations delivered by DRP~\cite{Topuz} are summarized in Table~\ref{DimenReduc} with respect to their geometrical classification. While spherical defects may be referred to the precipitates or voids, SFT are apparently the examples of tetrahedral objects. 
\begin{table} [H] 
\begin{center}
    \caption{Density and spacing formulations according to DRP~\cite{Topuz}}
   \begin{small}
    \begin{threeparttable}
     \begin{tabular}{*5l}
    \toprule
    Defect type & 2D density & 2D spacing & Line density & Line spacing  \\ 
      \midrule
    $\rm Spherical$ & $\frac{\pi^2}{3}\rho_{\rm Spheres}^{\rm 3D}R \tnote{*}$ & $ 1/\sqrt{\frac{\pi^2}{3}\rho_{\rm Spheres}^{\rm 3D}R}$ & $\frac{\pi^3}{6}\rho_{\rm Spheres}^{\rm 3D} R^{2}$ & $6/\pi^3 \rho_{\rm Spheres}^{\rm 3D} R^{2}$\\
      \midrule
    $\rm Tetrahedral$ & $\frac{\sqrt{6}}{3}\rho_{\rm Tetrahedra}^{\rm 3D}a \tnote{\dag}$ & $1/\sqrt{\frac{\sqrt{6}}{3}\rho_{\rm Tetrahedra}^{\rm 3D}a}$ &$\frac{\sqrt{6}}{6}\rho_{\rm Tetrahedra}^{\rm 3D}a^{2}$ & $\sqrt{6}/\rho_{\rm Tetrahedra}^{\rm 3D}a^{2}$ \\
     \midrule
    $\mbox{Prismatic loops}$ & $\rho_{\rm Loops}^{\rm 3D}d \tnote{\ddag}$ & $1/\sqrt{\rho_{\rm Loops}^{\rm 3D}d}$ & $\frac{1}{T \tnote{\S}}\sum\limits^{T \tnote{\S}}\frac{\sum\mbox{intersections}}{\sum\mbox{mesh length}}$ & $\left(\frac{1}{T}\sum\limits^{T}\frac{\sum\mbox{intersections}}{\sum\mbox{mesh length}}\right)^{-1}$ \\
   \bottomrule
    \end{tabular}
   \begin{tablenotes}
       \item[*] Average radius of spheres in 3D.
       \item[\dag] Average edge of regular tetrahedra in 3D.
       \item[\ddag] Average size of prismatic loops in 3D.
       \item[\S] Number of realizations.
   \end{tablenotes}
   \end{threeparttable}
    \label{DimenReduc}
     \end{small}
     \end{center}
    \end{table} 
Strength of the obstacles which determines pinning/depinning of edge dislocations is computed via DBH or BKS formulations depending on the nature of the corresponding defect. Substitution of the 2D spacing terms stated in Table~\ref{DimenReduc} into these formulations yields the expressions shown in Table~\ref{Strtab}.
\begin{table} [H] 
\begin{center}
    \caption{Strength formulations according to DRP~\cite{Topuz}}
    \begin{small}
     \begin{tabular}{*4c}
      \toprule
       & $\rm Spherical$ & $\rm Tetrahedral$ &  $\mbox{Prismatic loops}$   \\ 
      \midrule
    $\rm Strength$ & $A\mu b\left[\ln\left(\frac{\overline{D}}{r_{0}}\right)+B\right]\sqrt{\frac{\pi^2}{3}\rho_{\rm Spheres}^{\rm 3D}R}$  &  $0.67\mu b\sqrt{\frac{\sqrt{6}}{3}\rho_{\rm Tetrahedra}^{\rm 3D}a}$ & $0.33\mu b\sqrt{\rho_{\rm Loops}^{\rm 3D}d}$ \\
   \bottomrule
    \end{tabular}
     \label{Strtab}
     \end{small}
     \end{center}
    \end{table} 
If the parameters that constitute BKS formulation are examined, $A$ is a coefficient depending on the character of the dislocation, $A=1/2\pi(1-\nu)$ for screw dislocation and $A=1/2\pi$ for edge dislocation, $\nu$ is Poisson's ratio, $\mu$ is the shear modulus, $b$ is the Burgers vector magnitude, $B$  is a coefficient which is equal to 0.7 for precipitates and 1.52 for voids, $r_{0}$ is the line energy cut-off radius defining the elastic dislocation core size selected as $b$, and $\overline{D}$ is relative diameter defined as $\overline{D}=(D^{-1}+{L^{\rm 2D}_{\rm Discs}}^{-1})^{-1}$ where $L^{\rm 2D}_{\rm Discs}$ is the 2D spacing between the discs of spherical defects, and $D$ is the disc size taken as $2R^{\rm 2D}=4R/\pi$ since $R^{\rm 2D}=2R/\pi$ on average~\cite{Topuz}. In DBH model, the barrier strength coefficient, $\alpha$, which is scaled in the interval of [0.11, 1]~\cite{Lucas1993} is select as 0.67 and 0.33 for SFT and prismatic dislocation loops, respectively.
\section{Additional features for alloys}
\label{Additional features for alloys}
\subsection{Multi-type defects}
\begin{figure}[H]
\centering
\includegraphics[width=12cm]{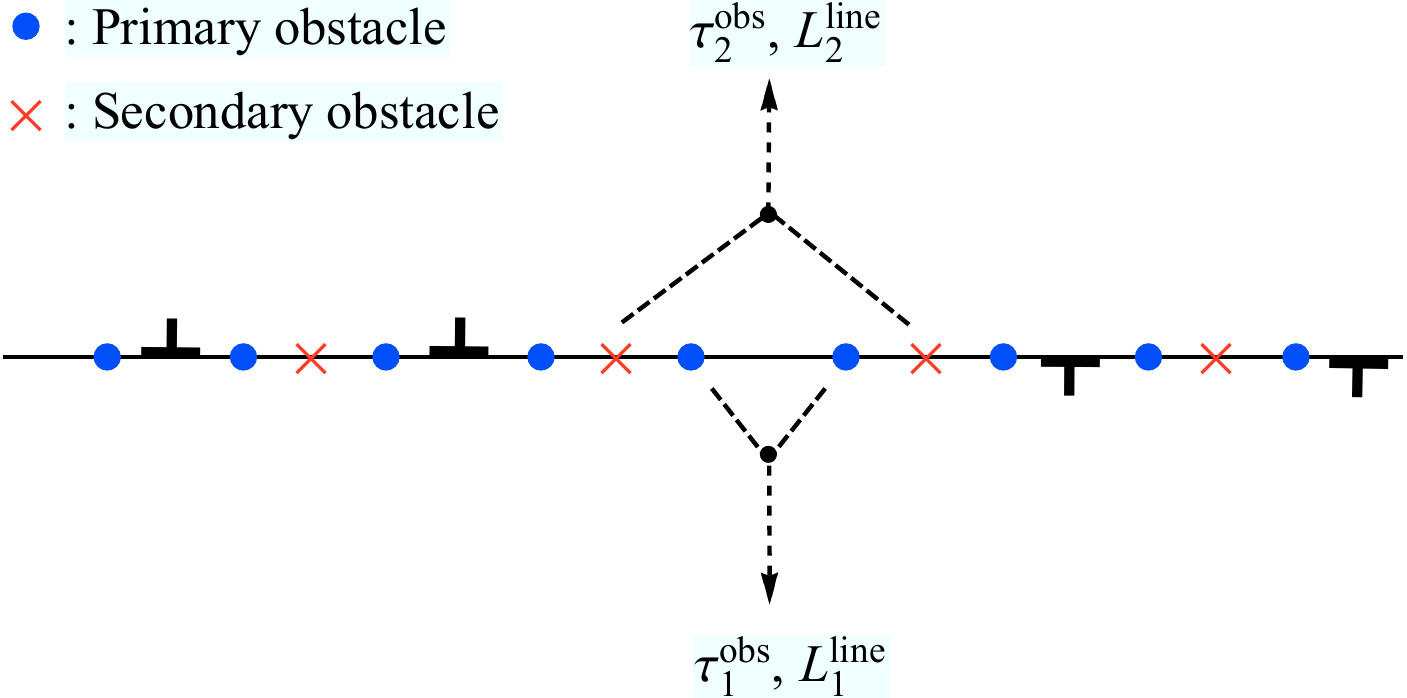}
\caption{Representation of multi-type defects according to their strength $\tau^{\rm obs}$ and line spacing $L^{\rm line}$ on slip lines.}
\label{multiype}
\end{figure}

In agreement with the experimental observations, the microstructure of alloys may not consist of only one type of defect. Depending on the nature of the process and the external load, different kinds of defects such as precipitates and dislocation loops may be present at the same time in the crystalline materials. Simultaneous existence of multi-type defects in the computational cell is the accurate description to show the condition of the material. From the perspective of DRP~\cite{Topuz}, each defect is characterized by a value for strength and line density on the slip line, and different levels of the triggering mechanisms such as irradiation or heat treatment are translated to a pair of strength and line density according to the changes in the 3D geometrical properties of the defects. Defining that $L^{\rm line}=1/\rho^{\rm line}$ where $L^{\rm line}$ is the line spacing between obstacles and  $\rho^{\rm line}$ is the line density, the distribution of multi-type obstacles is shown in Fig.~\ref{multiype}.
Precipitates may be considered as primary defects in alloys since their potency to change the yield point is generally superior in comparison with the other defects. In accordance with the crystal structure, SFT or dislocation loops may behave as secondary defects. It is worth to mention that this representation has no limitation for the number of defect type unless the total fraction is extremely high, and it is practical to enable the distribution of the multi-type defects as well as the sub-types of any category on slip lines.  

The combined effect of multi-type defects has been investigated in different studies~\cite{Brown1971, Lagerpusch2000, Hiratani2004, Monnet2006, Queyreau2010}, and the suggested resulting strength due to $i$ different defects is generally expressed by using the following expression:
\begin{equation}
\tau^{\rm obs}_{\rm res}=\left[\sum_{i}{\left[\tau^{\rm obs}_{i}\right]^{p}}\right]^{1/p}
\label{mixture}
\end{equation}
In spite of the existence of various values suggested for $p$ in the interval of [1,2] (e.g. $p$=1~\cite{Brown1971}), $p=2$ has been theoretically approved~\cite{Hiratani2004}, and it is experimentally observed~\cite{Lagerpusch2000}. It should be noted that line density, being an additional parameter in DRP for 2D DD, is as effective as strength in the case of single type defects~\cite{Topuz}, but it is not present in the mixture hardening rule stated in Eq.~\ref{mixture}. Another form of mixture law is the concentration-weighted superposition of two defect strengths:
\begin{equation}
\tau^{\rm obs}_{\rm res}=\left[(C_{1})^{q}(\tau^{\rm obs}_{1})^{p}+(C_{2})^{q}(\tau^{\rm obs}_{2})^{p}\right]^{1/p}
\label{mixconcen}
\end{equation}
where $C_{1}$ and $C_{2}$ are the concentration factors. $(p,q)=(1,0.5)$~\cite{Brown1971} and $(p,q)=(1,1$)~\cite{Kocks1980} are some of the proposed values for the pair of $p$ and $q$. A 3D DD study~\cite{Monnet2006} defines concentration factors by using the spacing between each type of obstacles, and suggests the following concentration based mixture law:
\begin{equation}
\tau^{\rm obs}_{\rm res}=\frac{l_{2}\tau^{\rm obs}_{1}+l_{1}\tau^{\rm obs}_{2}}{\sqrt{l_{1}^{2}+l_{2}^{2}}}
\label{mixconcenmon}
\end{equation}
where $l_{1}$ and $l_{2}$ are 2D spacing between obstacles of each type. According to DRP, line spacing $L^{\rm line}$ is associated with the concentration of the obstacles since line density determines the population of the obstacles on the slip lines. Therefore, one of the goals in the present study is to check if the behavior of yield strength under the condition of consecutive alignment described in Fig.~\ref{multiype} shows similarities with the mixture law in Eq.~\ref{mixconcenmon} when the concentration factors are defined in terms of line spacing $L^{\rm line}$.
\subsection{Internal friction}
Internal friction in 2D DD is the continuous resistance against the motion of edge dislocations on the slip lines. Obstacles are discrete objects on which dislocations pin/depin depending on the comparison between the strength of the obstacles and the absolute value of resolved shear stress $\tau$ of corresponding dislocations; however, internal friction in 2D DD is an implicit form of resistance that either sets dislocations immobile or influences the velocity of the dislocation by comparing the internal friction with $|\tau|$. The velocity term of an edge dislocation with a resolved shear stress $\tau$ affected by the internal friction denoted as $\tau_{f}$ gets the following form:
\begin{equation}
 v(\tau,\tau_{f}) =
  \begin{cases} 
   \hfill \frac{\left(\tau-\sgn(\tau)\tau_{f}\right)b}{B} \hfill &\text{if  $\lvert\tau\rvert > \tau_{f}$ }\\
   \hfill 0 \hfill &\text{ if  $\lvert\tau\rvert \leq \tau_{f}$ }
  \end{cases}
\label{fricvelo}
\end{equation}
where $B$ is the drag coefficient and $b$ is the Burgers vector magnitude. Whilst the internal friction may depend on some variables in the form of $\tau_{f}=\tau_{f}(x,y,t,T)$, which means it may be spatially changing $(x,y)$ or specific to any slip system, temporal $(t)$ or temperature-dependent $(T)$, present study assumes that $\tau_{f}$ is constant in the computation cell during any simulation for internal friction. The reference value  of internal friction is aimed to be the Peierls stress (PS) which is defined as the minimal stress to move a dislocation at 0 K~\cite{Kang2012}. An FCC lattice exerts a very weak resistance against the motion of dislocations: the PS lies within the order of $10^{-5}$ $\mu$ or less~\cite{Kubin1998}, where $\mu$ is the shear modulus; in contrast, PS is 900 MPa ($\approx1.1\times10^{-2}\mu$)~\cite{Marian2004} for screw dislocations in BCC Iron. Not only the PS value for the edge dislocation in the simulation material, but also values around this PS are investigated to depict the general behavior of yield strength under the effect of internal friction that slows down or immobilizes the edge dislocations.

\subsection{Dynamically increasing defect strength}
\begin{figure}[H]
\centering
\includegraphics[width=10cm]{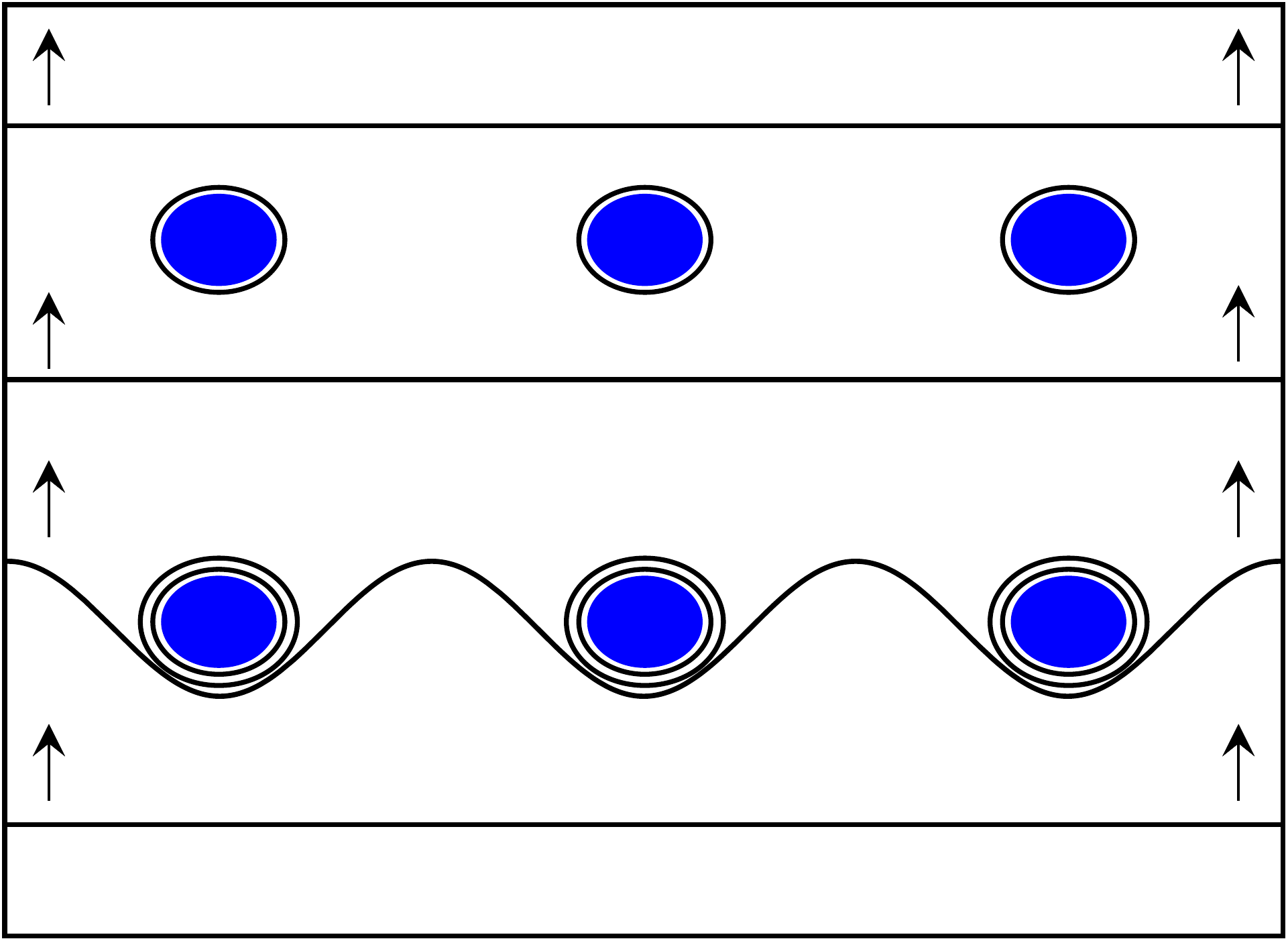}
\caption{Orowan loops on the precipitates.}
\label{OrowanLoop}
\end{figure}

\begin{figure}[H]
\centering
\includegraphics[width=13cm]{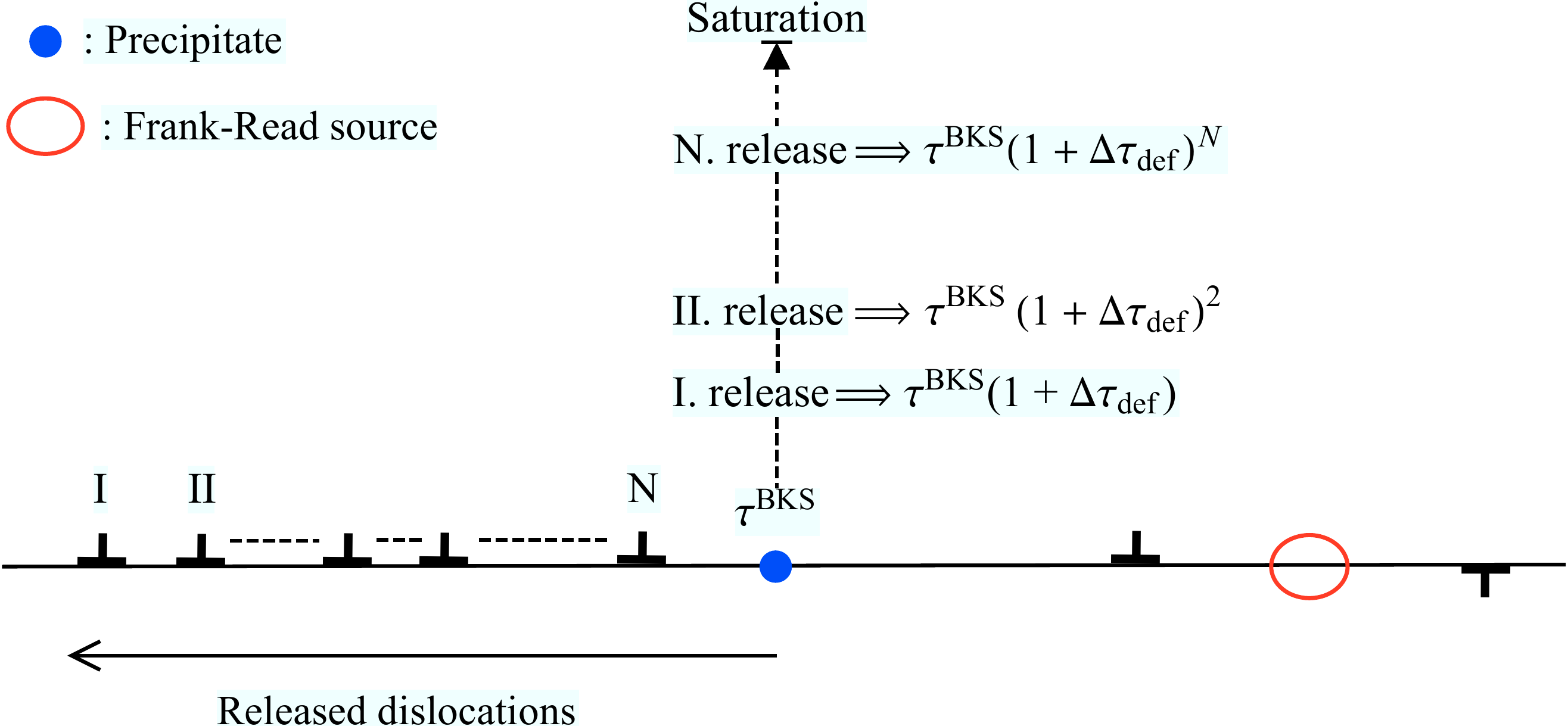}
\caption{Dynamically increasing defect strength.}
\label{dyninc}
\end{figure}

When a dislocation bypasses a row of particles, a so-called Orowan loop is produced after the release of this dislocation as illustrated in Fig.~\ref{OrowanLoop}. Orowan loops on the particles act to increase the required resolved shear stress of the next dislocation~\cite{Queyreau2010}; hence, loops together with particles behave as a combined barrier against the motion of the dislocations. 
In order to incorporate the effects of such a mechanism, a simple approach is applied using a recursive relation based on the number of dislocations released from precipitates. Specifying that $n$ is the number of released dislocations (i.e. the number of accumulated Orowan loops on an obstacle) and $\Delta\tau_{\rm def}$ is the percentage increase of the defect strength due to each Orowan loop, the strength of corresponding obstacle on which $(n+1)th$ edge dislocation is about to pin is hypothetically:
\begin{equation}
\tau_{n+1}^{\rm obs}=\tau_{n}^{\rm obs}(1+\Delta\tau_{\rm def}) \quad n=1,2,3...
\label{recursive}
\end{equation}
Since these obstacles are precipitates, and BKS model~\cite{BKS} is used to compute the initial strength of the precipitates, the relation between amplified strength due to the presence of $n$ loops denoted as $\tau_{n+1}^{\rm Inc}$ against the $(n+1)th$ interacting edge dislocation and initial strength $\tau^{\rm BKS}$ is expressed in a more generalized form:
\begin{equation}
\tau_{n+1}^{\rm Inc}=\tau^{\rm BKS}(1+\Delta\tau_{\rm def})^{n} \quad n=0,1,2...
\label{BKSrec}
\end{equation}
After the parametrization of dynamically increasing defect strength in terms of the initial value of the precipitate strength, the number of the released dislocations or accumulated loops, and a value for percentage increase, it is necessary to define a level of defect strength to terminate the algorithm. If there is such a value of saturation for strength, say $\tau^{\rm Sat}$, there also exists a maximum number of loops $N$ which determines the level of saturation according to Eq.~(\ref{BKSrec}):
\begin{equation}
\tau^{\rm BKS}(1+\Delta\tau_{\rm def})^{N}=\tau^{\rm Sat}
\label{SATdyninc}
\end{equation}
The progression of dynamically increasing precipitate strength is illustrated in Fig.~\ref{dyninc}. As it is hard to extract the value of the parameters in Eq.~(\ref{SATdyninc}) from experiments, the sensitivity of the increment value $\Delta\tau_{\rm def}$ is explored by solving Eq.~(\ref{SATdyninc}) for $N$, assuming that the ratio $\tau^{\rm Sat}/\tau^{\rm BKS}$ is known. 
To summarize, the strength of point obstacle referred to a precipitate is increased after each release of an edge dislocation from the corresponding barrier by imagining that an Orowan loop is formed, and this operation is terminated when the number of released dislocations reaches the saturation level which is indicated by the maximum number $N$.
\section{BCC slip systems and their effective Burgers vector magnitude}
\label{BCC slip systems and their effective Burgers vector magnitude}
\begin{figure}[H]
\centering
\includegraphics[width=12cm]{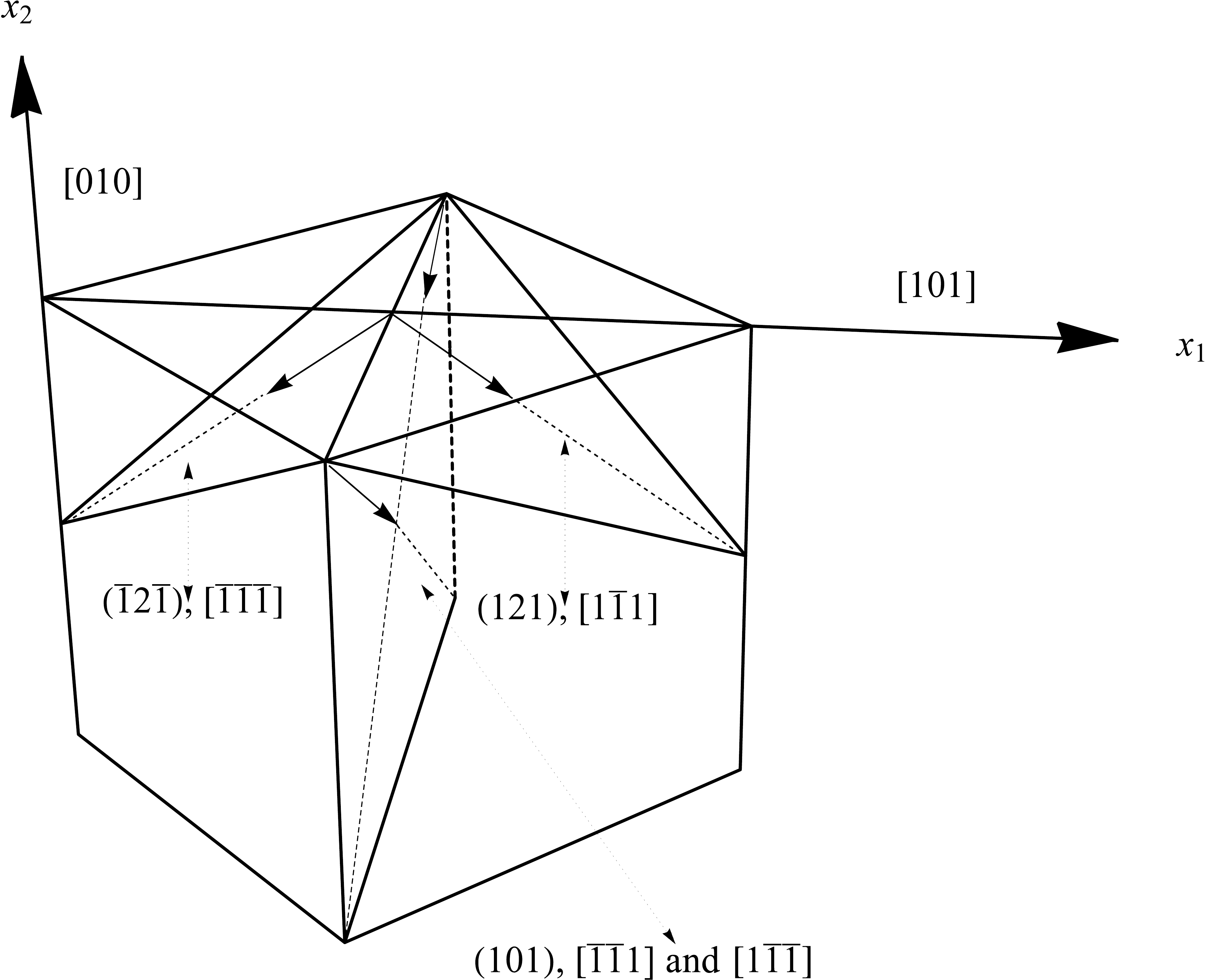}
\caption{Suggested~\cite{Ricebcc} slip systems for BCC crystal configuration.}
\label{Rice}
\end{figure}

The description of the plastic slip in a BCC crystal together with the suggested planes is originally taken from another study~\cite{Ricebcc} and it is given in Fig.~\ref{Rice}. In this description, the following slip planes are included: $(1 0 1)$, $(1 2 1)$, and $(\bar{1} 2 \bar{1})$ while the common slip direction in BCC is of $\langle1 1 1\rangle$ type.
\noindent
The Burgers vector magnitude of any member from $\langle1 1 1\rangle$ is:
\begin{equation}
|b|=\frac{a}{2}\sqrt{1+1+1}= \frac{a\sqrt{3}}{2}
\end{equation}
where $a$ is the size of the atomic unit cell.
\begin{figure}[H]
\centering
\includegraphics[width=8cm]{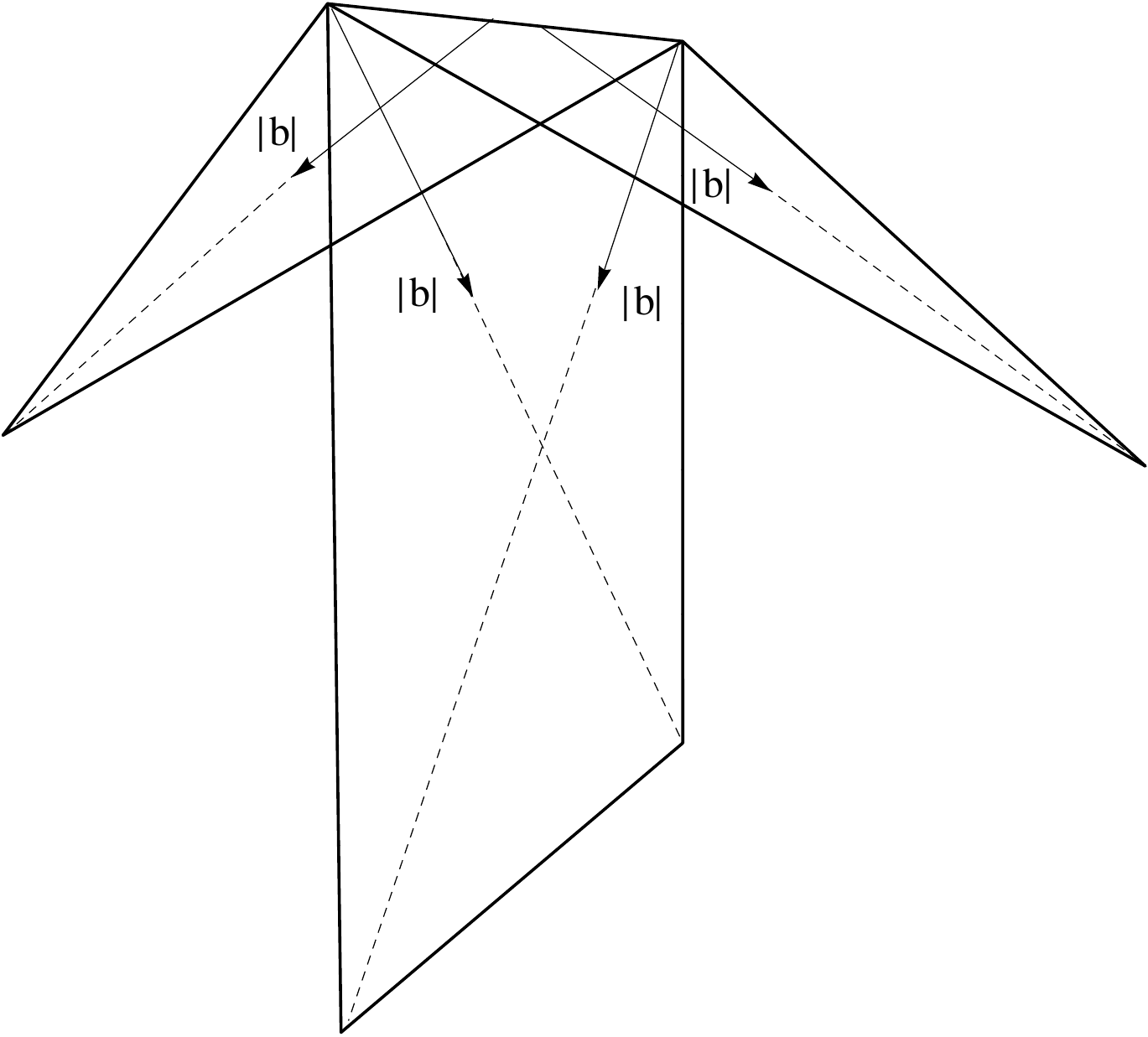}
\caption{Slip directions together with their Burgers vector magnitude in 3D.}
\label{3Db}
\end{figure}
One may determine the effective Burgers vector magnitude by projecting the corresponding vector on the trace of effective slip plane~\cite{Wang2009}. In order to accomplish this operation, Fig.~\ref{3Db} is separated into three planes. Considering that the edge of the cubic cell is $a$, $(1 0 1)$ is a rectangular plane whose edges are $a$ and $\sqrt{2}a$.
\begin{figure}[H]
\centering
\includegraphics[width=12cm]{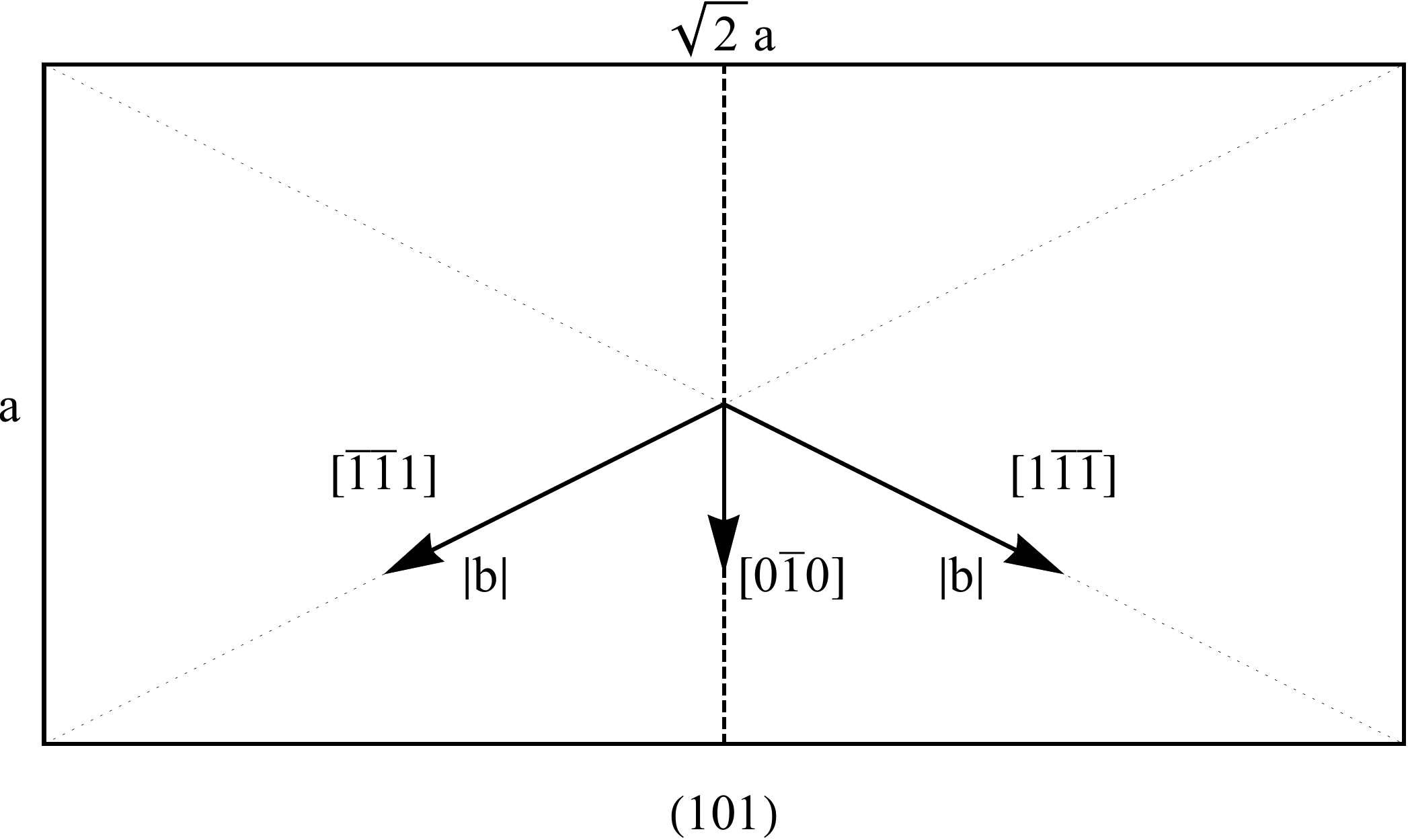}
\caption{$(1 0 1)$ plane with $[\bar{1}\bar{1}1]$ and $[1\bar{1}\bar{1}]$ directions.}
\label{101}
\end{figure}
The combination of equal slip on $(1 0 1)[\bar{1}\bar{1}1]$ and $(1 0 1)[1\bar{1}\bar{1}]$ yields an effective slip on $(1 0 1)[0\bar{1}0]$. The trace of the resulting slip direction is shown as the thick dashed line in Fig.~\ref{101}. Since both directions provide equal slip, projection of one direction onto the trace of effective direction is enough to calculate the effective Burgers vector magnitude. According to the geometry, $\cos([\bar{1}\bar{1}1]\angle[0\bar{1}0])=\sqrt{3}/3$. Hence, effective Burgers vector magnitude on $[0\bar{1}0]$ is $|b|\sqrt{3}/3$ where $|b|$ is the magnitude of Burgers vector for $[\bar{1}\bar{1}1]$ and $[1\bar{1}\bar{1}]$ directions.
\begin{figure}[H]
\centering
\includegraphics[width=8cm]{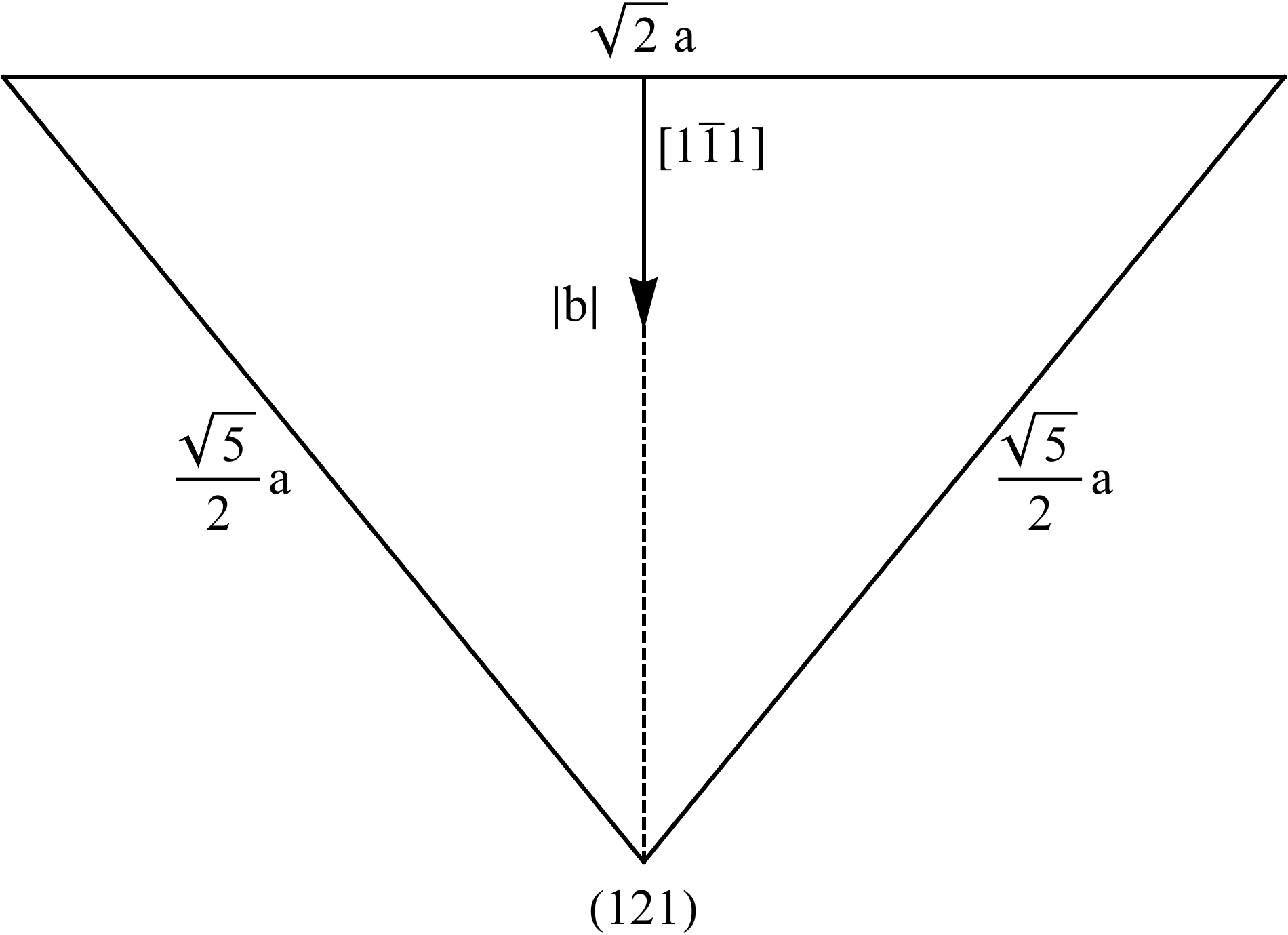}
\includegraphics[width=8cm]{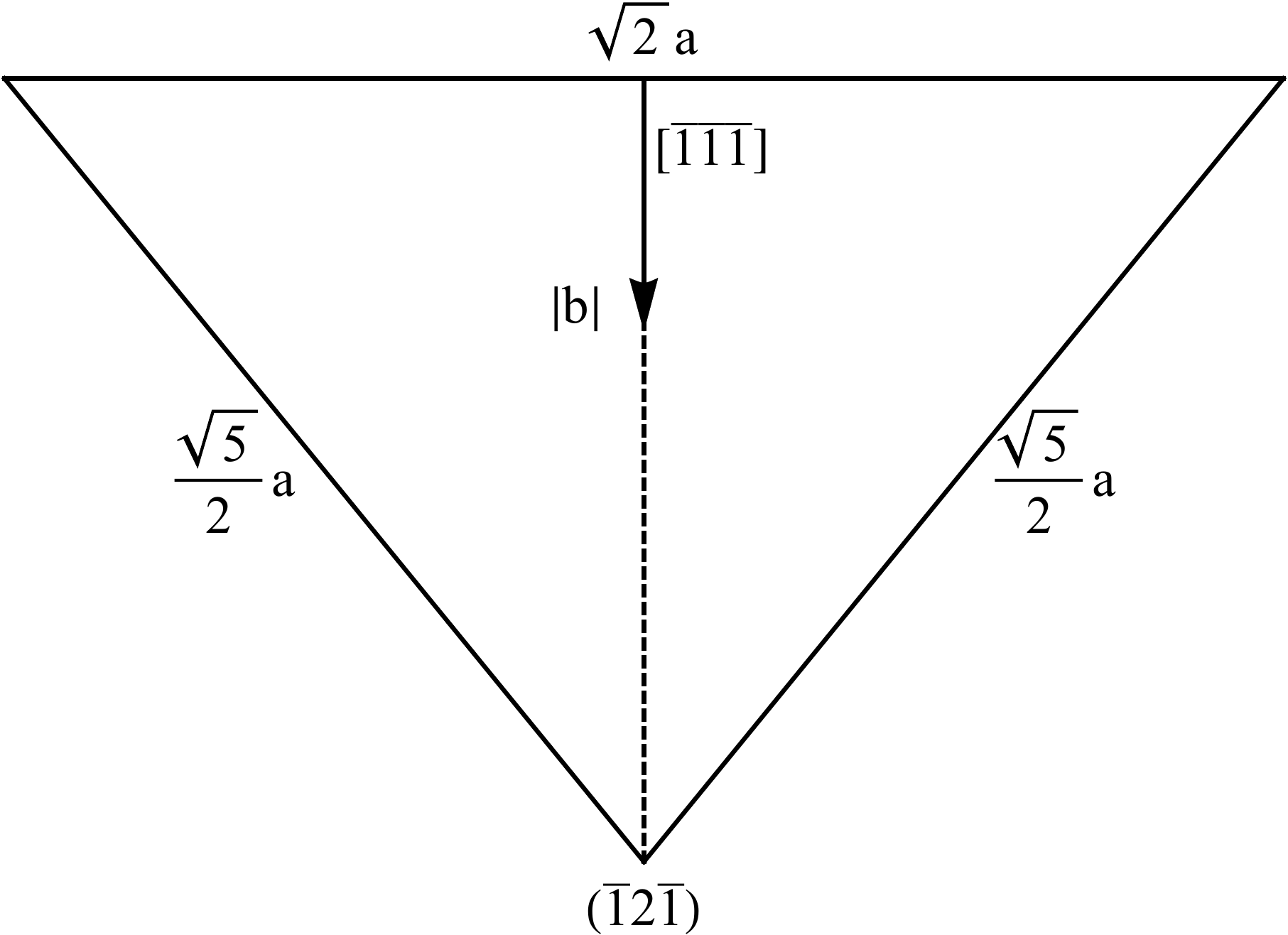}
\caption{$(1 2 1)[1\bar{1}1]$ and $(\bar{1} 2 \bar{1})[\bar{1}\bar{1}\bar{1}]$ slip systems.}
\label{121}
\end{figure}
$(1 2 1)$ and $(\bar{1} 2 \bar{1})$ are the remaining isosceles triangular planes with two edges of $a\sqrt{5}/2$ and one edge of $\sqrt{2}a$. Corresponding slip systems contribute individually to the global description. However, their contribution is similar since they are symmetrical around $[101]$. Fig.~\ref{121} shows their geometrical properties and Burgers vector direction. In both cases, Burgers vector direction in 3D coincides with the trace of the effective slip plane shown as thick dashed lines; therefore, slip direction and effective Burgers vector magnitude remain as identical.

\begin{figure}[H]
\centering
\includegraphics[width=9cm]{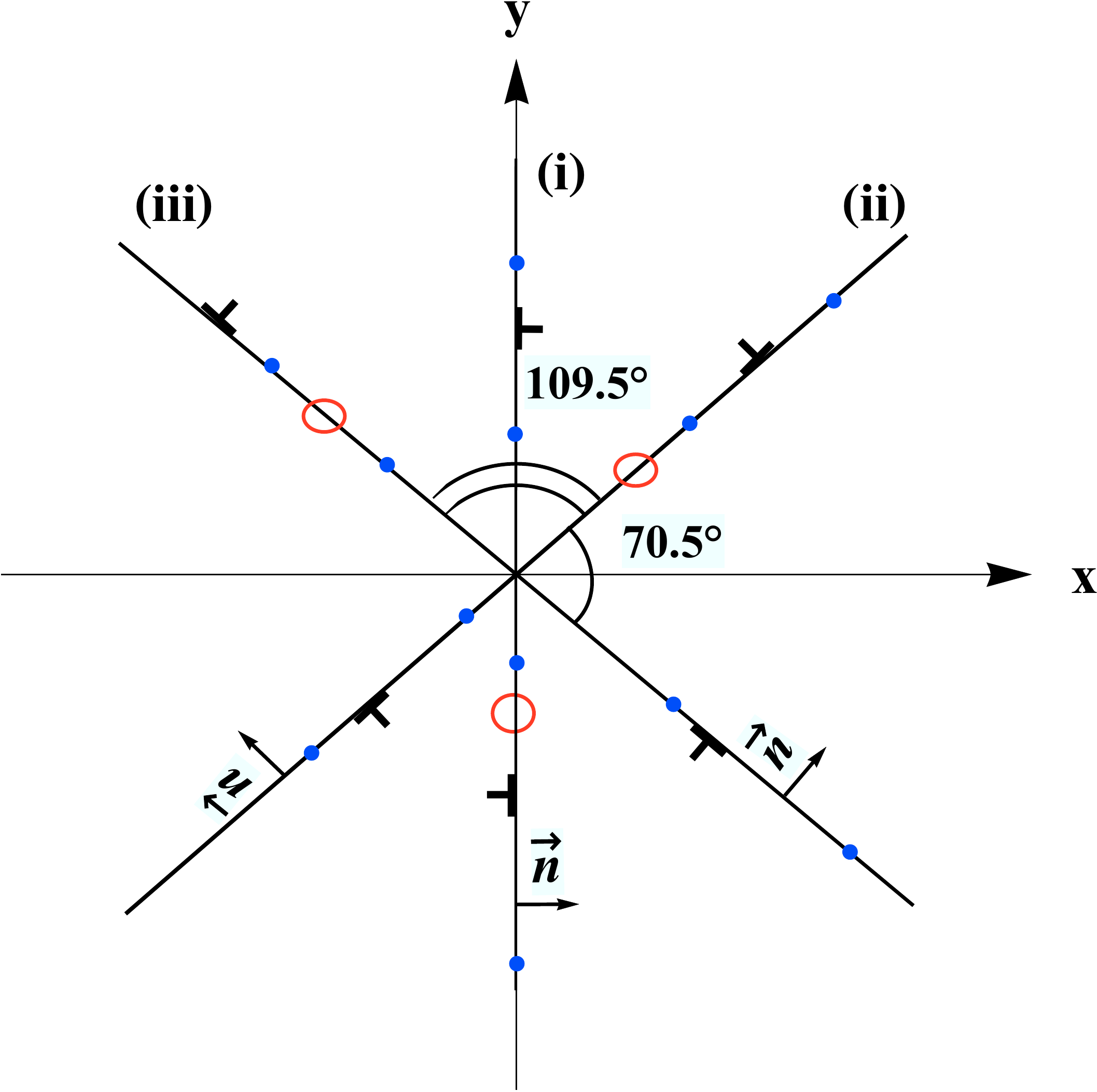}
\caption{Three slip systems with their inclination angles.}
\label{axes}
\end{figure}

To conclude, $[\bar{1}\bar{1}1]$ and $[1\bar{1}\bar{1}]$ on rectangular plane $(101)$ result in an effective slip direction on $(1 0 1)[0\bar{1}0]$ where the effective Burgers vector magnitude is $|b|\sqrt{3}/3$ in 2D. On triangular planes $(1 2 1)$ and $(\bar{1} 2 \bar{1})$, the slip directions $[\bar{1}\bar{1}\bar{1}]$ and $[1\bar{1}1]$ lie on the trace of the slip; hence, the effective Burgers vector magnitude is still $|b|$.These three slip systems in 2D are shown in Fig.~\ref{axes} and enumerated as follows:
\begin{enumerate}[(i)]
\item is $(1 0 1)[0\bar{1}0]$ with effective Burgers vector magnitude $|b|\sqrt{3}/3$;
\item is $(1 2 1)[1\bar{1}1]$ with effective Burgers vector magnitude $|b|$;
\item is $(\bar{1} 2 \bar{1})[\bar{1}\bar{1}\bar{1}]$ with effective Burgers vector magnitude $|b|$. 
\end{enumerate}
\section{Results and discussion}
\label{Results and discussion}
\subsection{Precipitate coarsening in CuCrZr}
The first material under investigation is CuCrZr~\cite{Edwards2005, CuCrZr} which is considered for the utilization in the first wall and divertor components of ITER~\cite{CuCrZr}. This material is subjected to certain processes such as prime aging (PA) and heat treatment (HT) for different periods of time in order to realize the modification in the precipitate properties. The reason of this operation is mentioned~\cite{CuCrZr} as the inability for the inhibition of dislocation motion during plastic deformation due to small size of precipitates~\cite{CuCrZr}, and annealing is applied on PA CuCrZr in order to coarsen the precipitates.

\begin{figure}[H]
\begin{center}
\includegraphics[width=14cm] {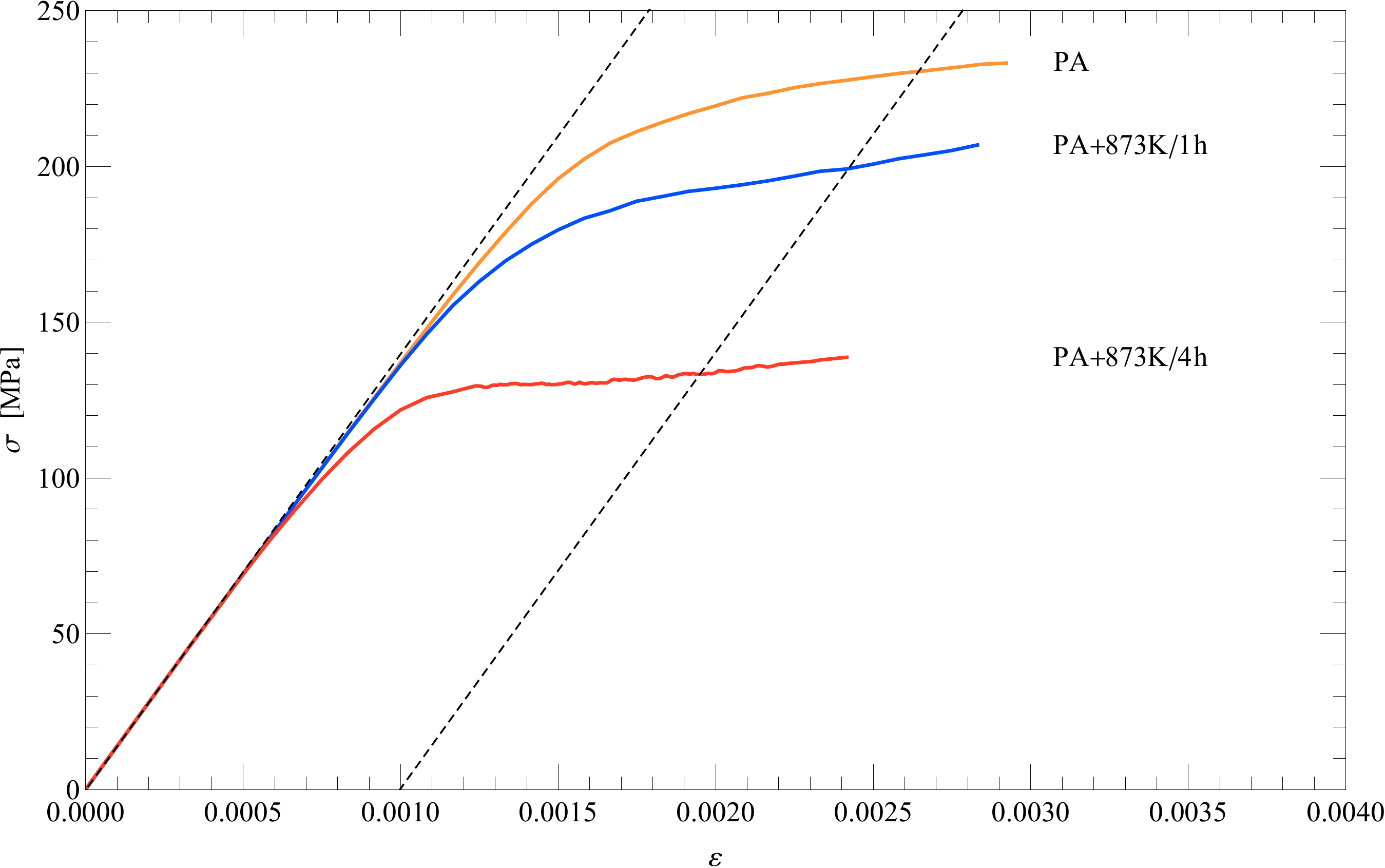}
\caption{Precipitate coarsening in single grain CuCrZr using constant defect strength.}
\label{ConsCuCrZr}
\end{center}
\end{figure}

Simulations are performed in a single grain of CuCrZr of size 30 $\rm \mu m$~\cite{Edwards2005} by using three cases, which are PA without HT, PA and HT for one hour (PA+873K/1h), and PA and HT for four hours (PA+873K/4h) at the temperature of 873 K. Common input parameters such as  elastic modulus $E$ and inter-planar distance $s$~\cite{Topuz}, which are used in all simulations of CuCrZr are given in Table~\ref{tabCuCrZr}. 
\begin{table} [H] 
\begin{center}
    \caption{Common input in all CuCrZr simulations}
    \begin{tabular}{*7c}
    \toprule
    $E$ (GPa) & $\nu$ & $b$ (nm)& Strain rate ($\rm s^{-1}$) & $s$ & Grain size  ($\mu \rm m $) & FR density ($\mu \rm m^{-2}$) \\
    \midrule
    123 & 0.34 & 0.25 & 2500 & 200$b$ & 30  & 1 \\
    \bottomrule
    \end{tabular}
    \label{tabCuCrZr}
\end{center}
    \end{table}
Initially, DRP is applied to the precipitates of CuCrZr by using the size (i.e. diameter) and density data delivered from an experimental study~\cite{CuCrZr}. Table~\ref{DRCuCrZr} presents the experimental data together with the line density and strength values obtained via dimension reduction.
\begin{table} [H] 
\begin{center}
\caption{Precipitate properties in CuCrZr}
\begin{tabular}{*5c}
\toprule
Process & Size  $(\rm nm)$ & 3D density $(\mu \rm m^{-3})$ & Line density $(\mu \rm m^{-1})$ & Strength (MPa)  \\
\midrule 
PA & 2.2 & $ 2.6\times10^{5}$ & 1.62 & 134 \\
PA+873K/1h & 8.7 & $0.17\times10^{5}$ & 1.66 & 106\\
PA+873K/4h & 21.3 &  $0.015\times10^{5}$ & 0.88 & 61\\
\bottomrule
\end{tabular}
\label{DRCuCrZr}
\end{center}
\end{table}
When the size and density values of the precipitates in Table~\ref{DRCuCrZr} are examined, it is clear that the size of precipitates increases significantly as a consequence of the HT; in contrast, this process results in the sharp decrease of density values. Stress-strain curves obtained by the application of DRP and static precipitate strength are shown in Fig.~\ref{ConsCuCrZr}. 
It is seen that the yield strength of the material is led by properties of the precipitates and it is lowered by the coarsening due to HT. According to Fig.~\ref{ConsCuCrZr}, PA+873K/4h shows a more significant yield drop in comparison with PA+873K/1h. Whereas the value of the line density in PA+873K/1h indicates an inconsequential change with respect to PA, PA+873K/1h has a remarkably lower number of defect points than both PA and PA+873K/1h cases. The strength of precipitates is reduced in PA+873K/1h as well as PA+873K/4h, but the amount of decline is higher in PA+873K/4h. The stress-strain curves shown in the experimental study~\cite{CuCrZr} differs from Fig.~\ref{ConsCuCrZr} in the position of  PA+873K/1h. According to these experimental tensile results~\cite{CuCrZr}, the gap between the yield strength of PA and PA+873K/1h is large; on the other hand, the difference between PA+873K/4h and  PA+873K/1h is much smaller than this gap. There may exist several causes for this dissimilarity, nevertheless different perspectives may interpret this difference in a variety of ways. For instance, it is prevalent to associate the defect strength with the yield strength of the material by means of a linear relation between the change in yield strength and defect strength. Even if the line density, which is a fundamental parameter in the DRP is ignored, the defect strengths shown in Table~\ref{DRCuCrZr} do not support the experimental proximity of PA+873K/1h to PA+873K/4h. On the contrary, defect strength values computed via BKS model imply that the gap between PA+873/1h and PA+873K/4h may be approximately 1.6 times greater than the difference between PA and PA+873/1h. Over and above, formulations for the estimations of defect strength are not unique~\cite{DBH, BKS, Monnet2006, Queyreau2010}, and contradicting as well as supporting formulations may be found for this specific case, but they may 	
reverse their roles in another case. 

\begin{figure}[H]
\begin{center}
\includegraphics[width=14cm]{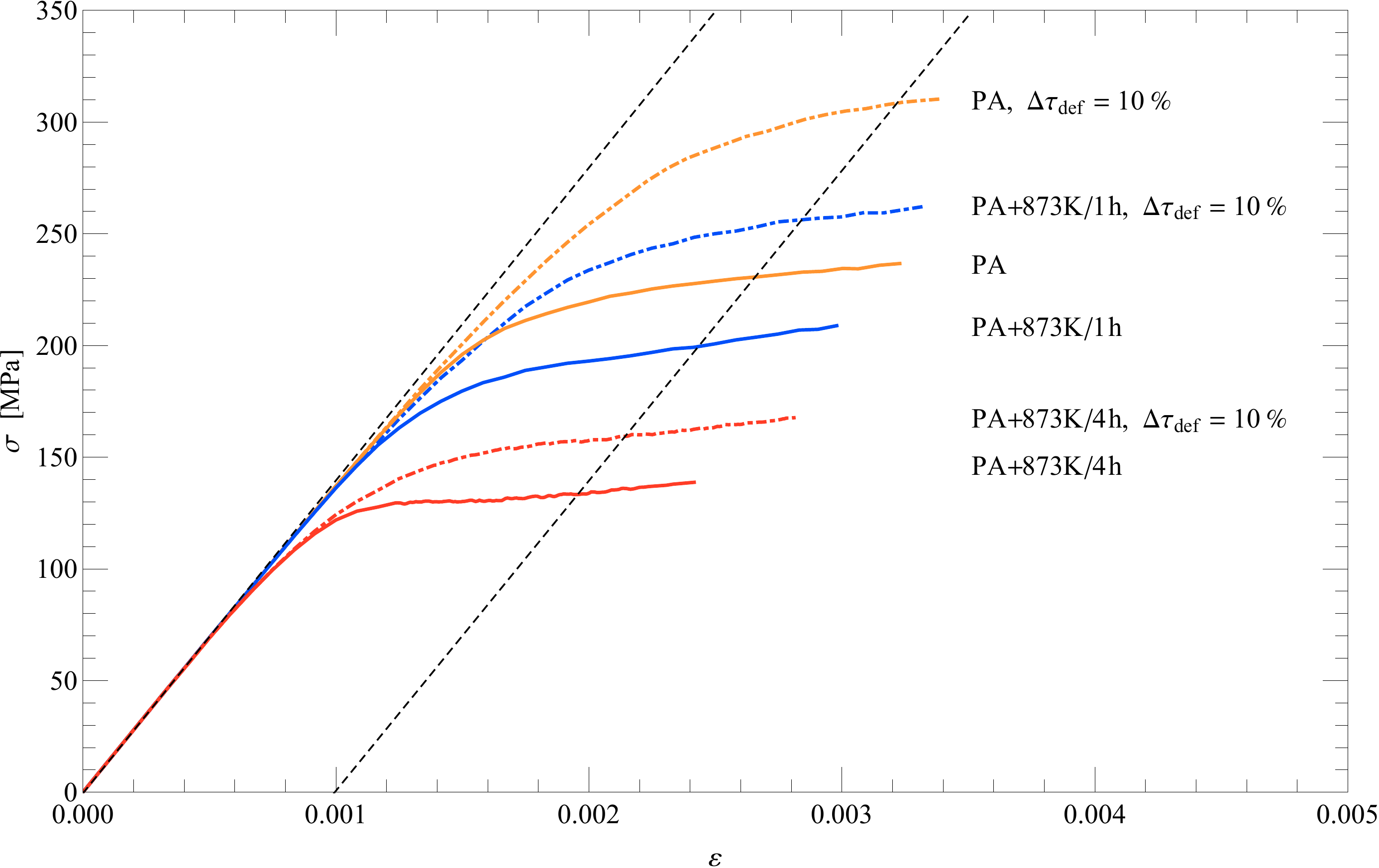}
\caption{Dynamically increasing precipitate strength in single grain CuCrZr with the use of defect strength increment 10$\%$ and saturation ratio 2.}
\label{BKSvs10INC}
\end{center}
\end{figure}

\begin{figure}[H]
\begin{center}
\includegraphics[width=14cm]{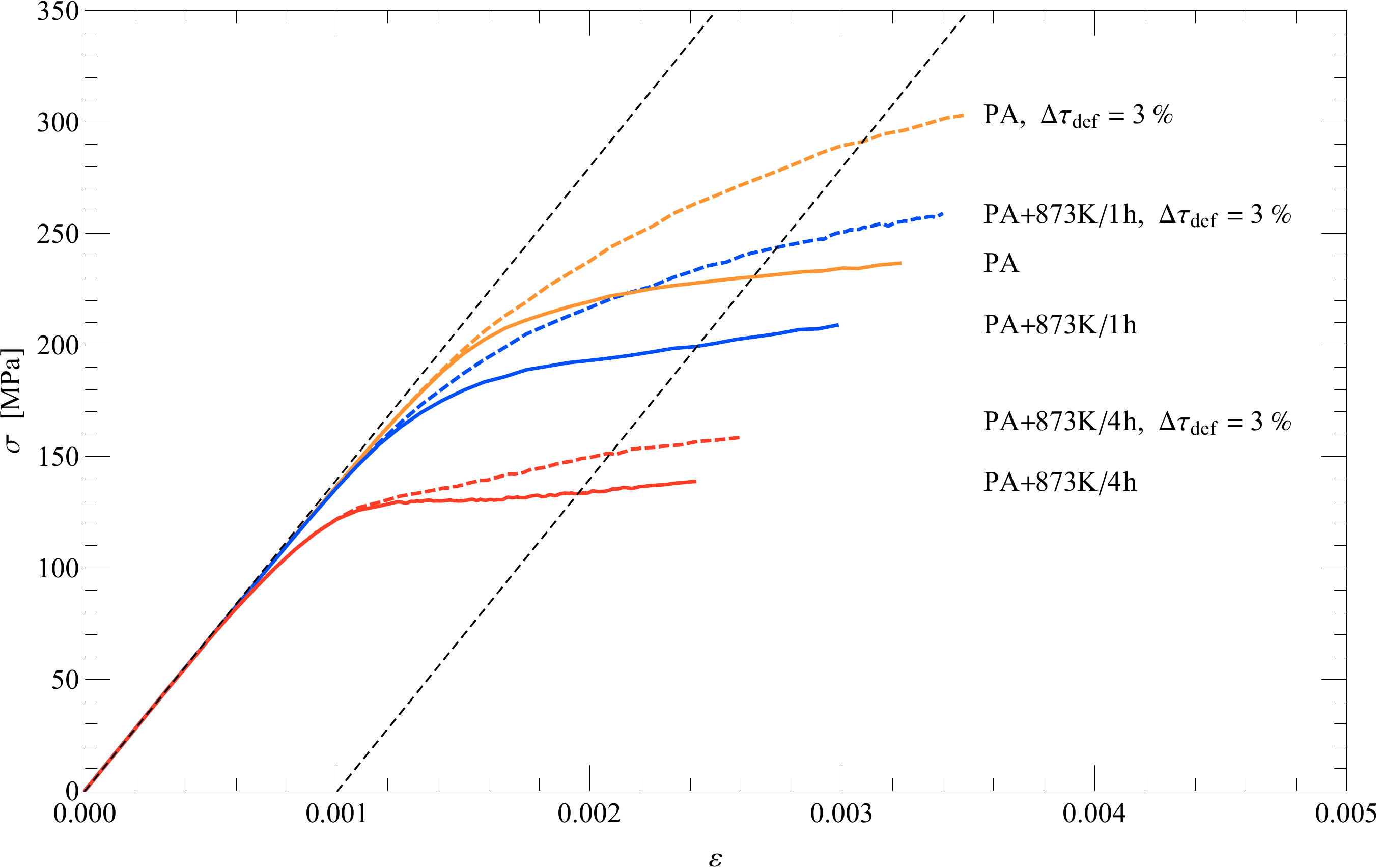}
\caption{Dynamically increasing precipitate strength in single grain CuCrZr with the use of defect strength increment 3$\%$ and saturation ratio 2.}
\label{BKSvs3INC}
\end{center}
\end{figure}

Dynamically increasing defect strength is also applied on CuCrZr by using the same precipitate properties mentioned in Table~\ref{DRCuCrZr}. The parameters of the first trial are inspired from another study~\cite{Queyreau2010} dedicated to the Orowan mechanism. The ratio between the saturation level and the initial strength value $\tau^{\rm Sat}/\tau^{\rm BKS}$ is assumed to be 2, and $\Delta\tau_{\rm def}$ is set to be 10$\%$. When these inputs are substituted into  Eq.~(\ref{SATdyninc}), the maximum number of loops $N$ after which the process of increase is terminated is found to be around 7. The result of this simulation is given in Fig.~\ref{BKSvs10INC}.
The dynamic increase of precipitate strength causes a significant rise in the yield strength, and the contribution of this feature proportionally increases with respect to precipitate properties as observed in Fig.~\ref{BKSvs10INC}. A remarkable property of this case is that the use of $\Delta\tau_{\rm def}=10\%$ together with $\tau^{\rm Sat}/\tau^{\rm BKS}=2$ does not change the behavior of stress-strain curves sharply except a shift in the yield strength. In order to test the effect of increment value, $\Delta\tau_{\rm def}=3\%$ is employed with the same ratio, for which the number of dislocation releases, $N$, needed to terminate the strengthening process is calculated as 23. The simulation output is illustrated in Fig.~\ref{BKSvs3INC}. It is observed that when $\Delta\tau_{\rm def}=3\%$ is applied instead of 10$\%$, the hardening characteristic is distinct, and the 0.1$\%$ offset yield point gets lower values when compared with the previous case. In short, according to both Fig.~\ref{BKSvs10INC} and Fig.~\ref{BKSvs3INC}, augmentation of precipitate strength based on dislocation release shows notable effect on the variation of yield strength.
\subsection{Internal friction in Ta-2.5W}
\begin{figure}[H]
\begin{center}
\includegraphics[width=14cm]{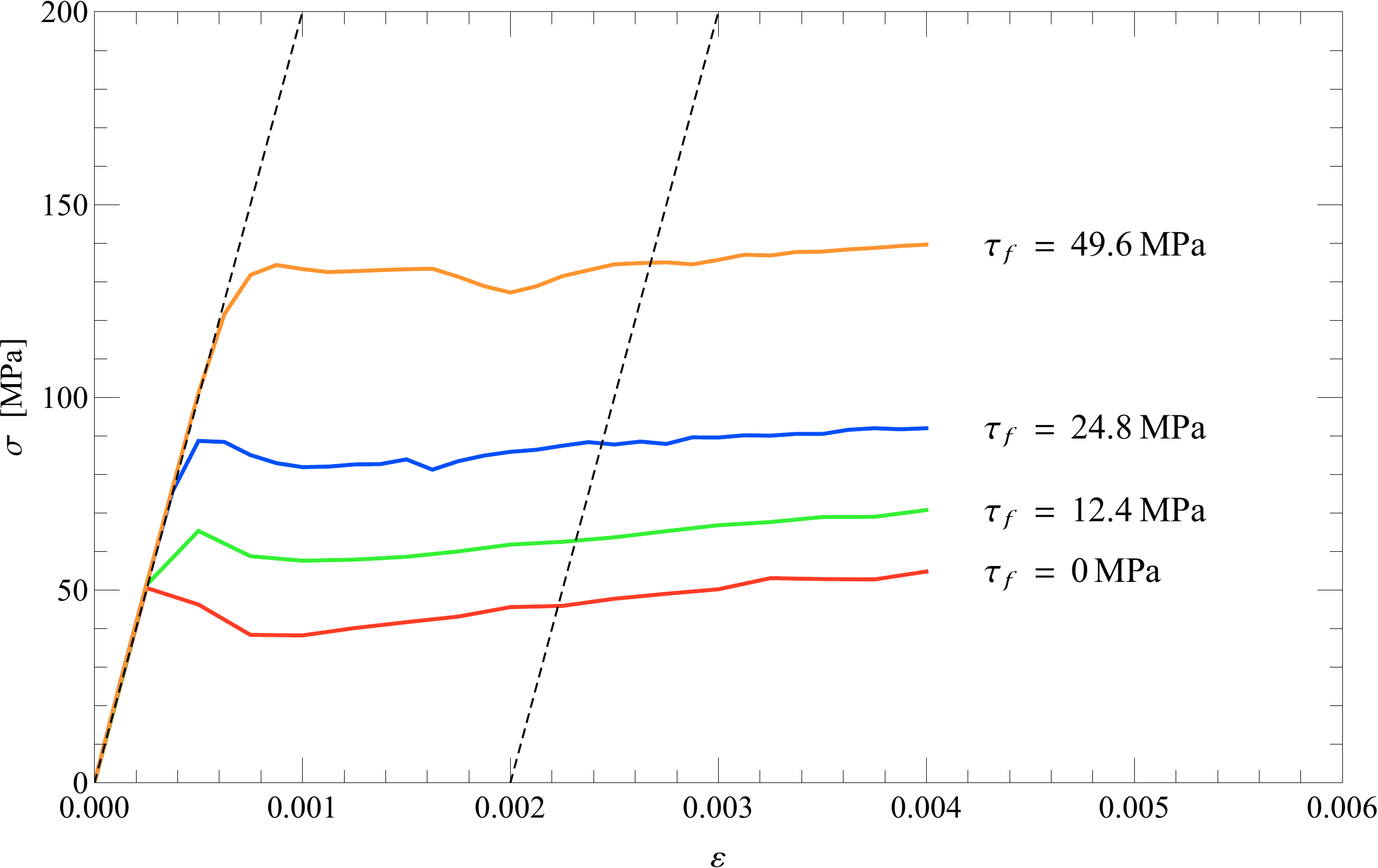}
\caption{Internal friction $\tau_{f}$ in single grain Ta-2.5W.}
\label{FRICTa}
\end{center}
\end{figure}

Tantalum and its alloys are defined as refractory metals thanks to their excellent resistance to corrosion and heat. These beneficial properties make them preferable also in nuclear applications~\cite{Byun2008}. Therefore, single grain Ta-2.5W alloy of grain size 40 $\rm \mu m$  (taken instead of 45 $\rm \mu m$ ~\cite{Khan1999}) is selected for the simulations of internal friction which are performed by the implementation of friction-dependent mobility law stated in Eq.~(\ref{fricvelo}). The reference point is decided to be  24.8 MPa which is the PS value used in a 3D DD study~\cite{Wang2011} for edge dislocations in Ta. Not only this PS value, but also $0.5\rm PS$  and $2\rm PS$ are checked to comprehend the contribution of friction values on slip lines. Additionally, BCC slip systems and their effective Burgers vector magnitudes stated in section~\ref{BCC slip systems and their effective Burgers vector magnitude} are also incorporated for the first time. The input of the simulations for Ta-2.5W is gıven in Table~\ref{Tainput}.
\begin{table} [H]
\begin{center}
    \caption{Basic input for Ta-2.5W simulations}
    \begin{tabular}{*7c}
    \toprule
    $E$ (GPa) & $\nu$ & $b$ (nm) & Strain rate ($\rm s^{-1}$) & $s$ & Grain size  ($\mu \rm m $)  &  FR density ($\mu \rm m^{-2}$) \\
    \midrule
    186 & 0.34 & 0.25 & 2500 & 400$b$ & 40 & 1 \\
    \bottomrule
    \end{tabular}
    \label{Tainput}
   \end{center}
    \end{table}
The results obtained for three non-zero friction values including the PS value are sketched in Fig.~\ref{FRICTa}, and it is demonstrated that all $\tau_{f}$ values result in parallel shifts in plastic response. Furthermore, when the yield points are determined by plotting the offset at 0.2$\%$ plastic strain, $\sigma_{0.2}=\{45.9, 62.7, 88.7, 134.7\}$ is acquired for $\tau_{f}=\{0, 12.4, 24.8, 49.6\}$. Defining that $\sigma_{0}$ is the $\sigma_{0.2}$ of frictionless case, the fraction $(\sigma_{0.2}-\sigma_{0})/\tau_{f}$ yields 1.35, 1.72, and 1.79 for $\tau_{f}\neq0$. Thus, the contribution of internal friction is not entirely linear. However, a coarse approximation such that $\sigma_{0.2}\approx\sigma_{0}+\beta\tau_{f}$, where $\beta$ is around 1.75 for an FR density of 1  $\mu \rm m^{-2}$ in Ta-2.5W, may be obtained for this specific case. The PS value causes a very considerable rise in the yield strength by an amount of 93$\%$. The simulations done for internal friction explicitly indicate that friction is an effective term which shifts the yield strength of the material, hence the presence/absence of friction in the mobility laws makes a significant difference. However, it is not always straightforward to predict the values of friction, and it may require a multi-scale approach.
\subsection{Mixed hardening in irradiated ODS EUROFER with 0.3$\%$ yttria}
Ferritic/martensitic alloys, especially ODS steels, have been taken into account as candidate materials~\cite{Heintze2011} for the components of nuclear systems such as generation-IV as well as fusion reactors thanks to their resistance occasioned by the existence of highly dispersed oxide particles to swelling and low damage accumulation~\cite{Schaublin2006}. For this reason, polycrystalline ODS EUROFER of grain size 5 $\mu \rm m$~\cite{Lindau2002, Eiselt2009} with 0.3$\%$ yttria in which  constituents are 8.9 wt$\%$ Cr, 1.1 wt$\%$ W, 0.47 wt$\%$ Mn, 0.2 wt$\%$ V, 0.14 wt$\%$ Ta, 0.11 wt$\%$ C and Fe for the balance is selected for the simulations of mixed hardening. ODS steel samples are irradiated with 590 MeV protons to 0.3, 1, and 2 dpa, and the properties of the irradiation-induced defects are chosen by using the observations of transmission electron microscopy (TEM) in an experimental study~\cite{Ramar2007}. The dispersion properties in unirradiated material are included from another experimental study~\cite{Heintze2011} where small angle neutron scattering (SANS) is the technique to investigate the particle properties. In both experiments, samples are provided from Plansee~\cite{Ramar2007, Heintze2011}. The reason of this combination is the possible variation of dispersion properties with the initiation of irradiation, since another experimental method reports concretely different results, and there exists a remarkable rise in the ultimate tensile strength of the material~\cite{Schaublin2006}. 
\begin{figure}[H]
\begin{center}
\includegraphics[width=16cm]{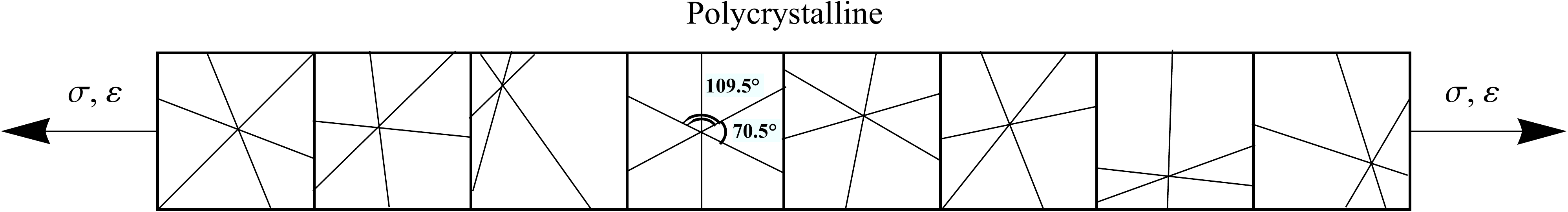}
\caption{Polycrystalline ODS EUROFER with 0.3$\%$ yttria for plain strain deformation.}
\label{PolyODS}
\end{center}
\end{figure}
The polycrystalline computational cell of ODS EUROFER with 0.3$\%$ yttria for plain strain deformation is shown in Fig.~\ref{PolyODS}. The polycrystalline material is composed of eight grains, and the size of each grain is 5 $\mu \rm m$. Slip systems are randomly generated by conserving the angles between slip lines, and effective Burgers vector magnitudes are used. The input parameters of this simulation are given in Table~\ref{inputodseuro}.
\begin{table} [H] 
\begin{center}
    \caption{Basic input for ODS EUROFER with 0.3$\%$ yttria simulations}
    \begin{tabular}{*7c}
    \toprule
    $E$ (GPa) & $\nu$ & $b$ (nm)& Strain rate ($\rm s^{-1}$) & $s$ & Grain size  ($\mu \rm m $)  &  FR density ($\mu \rm m^{-2}$) \\
    \midrule
    211.7 & 0.34 & 0.286 & 2500 & 200$b$ & 5 & 5 \\
    \bottomrule
    \end{tabular}
    \label{inputodseuro}
       \end{center}
        \end{table}
As it is mentioned above, the precipitate properties of the unirradiated material are taken from an experimental study with SANS, and the defect properties including both precipitates and irradiation induced defects of the irradiated material are found in another experimental work where the investigation technique is TEM. All irradiation-induced defects are assumed to be sessile prismatic loops in addition to yttria particles. The experimental data about defect properties in unirradiated and irradiated cases are presented in Table~\ref{tabodseuro}.
\setlength{\tabcolsep}{2pt}
\begin{table} [H] 
\begin{center}
    \caption{3D defect properties in unirradiated~\cite{Heintze2011} and irradiated~\cite{Ramar2007} case}
    \begin{tabular}{*5c }
    \toprule
    Irra. dose (dpa) & Disper. size (nm) & Disper. density ($\mu \rm m^{-3}$) & Loop size (nm) & Loop density ($\mu \rm m^{-3}$) \\ 
    \midrule
    0 & 3.8 & $11.5\times10^{4}$ & - & - \\  
    0.3 & $6-10$ & $4.5\times10^{4}$ & $1-2$ & $2.3\times10^{4}$ \\  
    1 & $6-10$ & $4.5\times10^{4}$ & $2-3$ & $4.4\times10^{4}$ \\  
    2 & $6-10$ & $4.5\times10^{4}$ & $5-10$ & $5.1\times10^{4}$ \\ 
    \bottomrule
 \end{tabular}
    \label{tabodseuro}
     \end{center}
     \end{table}
DRP is applied on the properties of defects mentioned in Table~\ref{tabodseuro} in accordance with the formulations stated in Table~\ref{DimenReduc}. BKS model is used to calculate the strength of precipitates whereas the strength of prismatic dislocation loops is determined by DBH model together with $\alpha=0.33$~\cite{Odette1979}.
\begin{table} [H] 
\begin{center}
\caption{Line density and defect strength according to the data in Table~\ref{tabodseuro}}
\begin{tabular}{*5c }
\toprule
{} &  \multicolumn{2}{c}{Dispersion}  & \multicolumn{2}{c}{Prismatic loop}\\ 
\midrule
Irra. dose (dpa) & Line density ($\mu\rm  m^{-1}$) & Strength (MPa) & Line density ($\mu \rm m^{-1}$) & Strength (MPa) \\
\midrule 
0 & 2.14 & 268 & - & - \\
0.3 & 3.72 & 303 & 0.03 & 44\\
1 & 3.72 & 303 & 0.17 & 78\\
2 & 3.72 & 303 & 1.74 & 146\\
\bottomrule
\end{tabular}
\label{ldnstreuro}
\end{center}
\end{table}
A comparison of line density and strength values in Table~\ref{ldnstreuro} aids to predict the yield behavior of the present material. Starting with the unirradiated case, the yield strength of four cases are labeled as $\sigma_{0}$, $\sigma_{0.3}$, $\sigma_{1}$, and $\sigma_{2}$ respectively. When the dispersion data of unirradiated material is compared with the case of 0.3 dpa, it is seen that the line density changes significantly, whereas the strength of the dispersions varies slightly. This change guarantees hardening; thus, $\sigma_{0.3}>\sigma_{0}$. The contribution of dislocation loops at this level is inactive due to very low line density and weak strength. For the next irradiation level, 1 dpa, there is no change in the dispersion data. A remarkable difference in the properties of irradiation induced defects is observed; however, line density is still very low. Therefore, one may expect that $\sigma_{1}\approx\sigma_{0.3}$. For the final irradiation level, besides the same dispersion properties, it is expected that irradiation induced defects result in a weak but visible hardening since almost every slip line has one or more interactions with these loops and the strength of the defects is far from the strength of FR sources. To conclude, yield strength of these four cases can be sorted as follows: $\sigma_{0}<\sigma_{0.3}\approx\sigma_{1}<\sigma_{2}$.
\begin{figure}[H]
\begin{center}
\includegraphics[width=14cm]{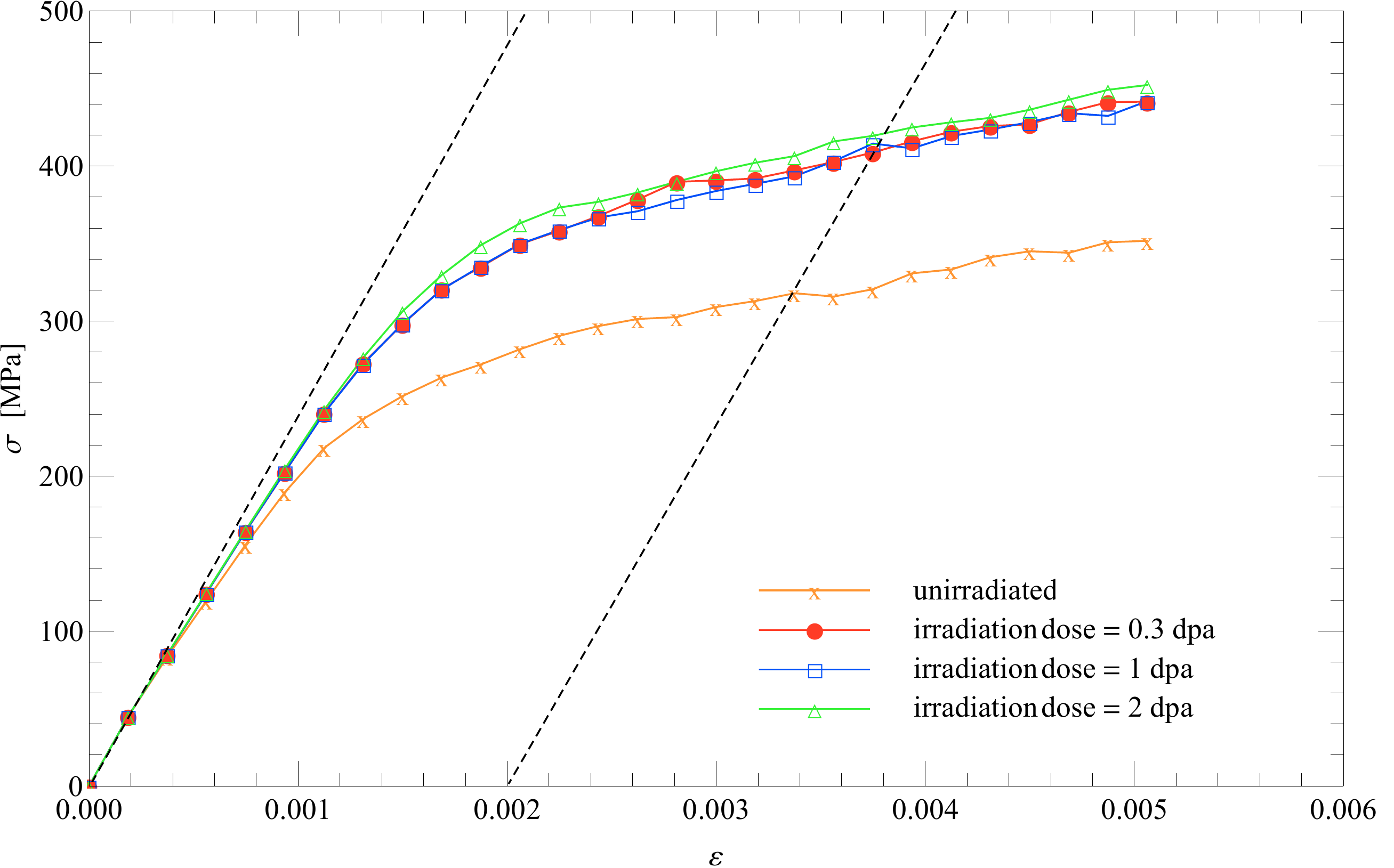}
\caption{Plain strain response of polycrystalline ODS EUROFER with 0.3$\%$ yttria using Tables~\ref{inputodseuro} and~\ref{ldnstreuro}.}
\label{ODSEURODS5}
\end{center}
\end{figure}
The simulation results reported in Fig.~\ref{ODSEURODS5} confirm the initial expectations stated above, and dislocation loops make a weak contribution to hardening. The outcome of these simulations is also qualitatively consistent with the concentration based mixture law stated in Eq.\ref{mixconcenmon} if the concentration factors are expressed in terms of line spacing $L^{\rm line}$. When the secondary defect is much weaker and less populated than the primary defect, the resulting defect strength is not far from that of the primary defect. Consequently, the yield strength is principally governed by the strength of the primary defect.

In order to see the effect of stronger secondary defects, an artificial case may be imagined by considering that barrier strength coefficient $\alpha$ in the strength formulation of prismatic loops stated in Table~\ref{Strtab} may get greater values than 0.33. Therefore, the final simulation is dedicated for the investigation of $\alpha=\{0.33, 045, 0.57, 0.68\}$ by the use of geometrical properties stated for irradiation dose of 2 dpa in Table~\ref{tabodseuro}.
\begin{figure}[H]
\begin{center}
\includegraphics[width=14cm]{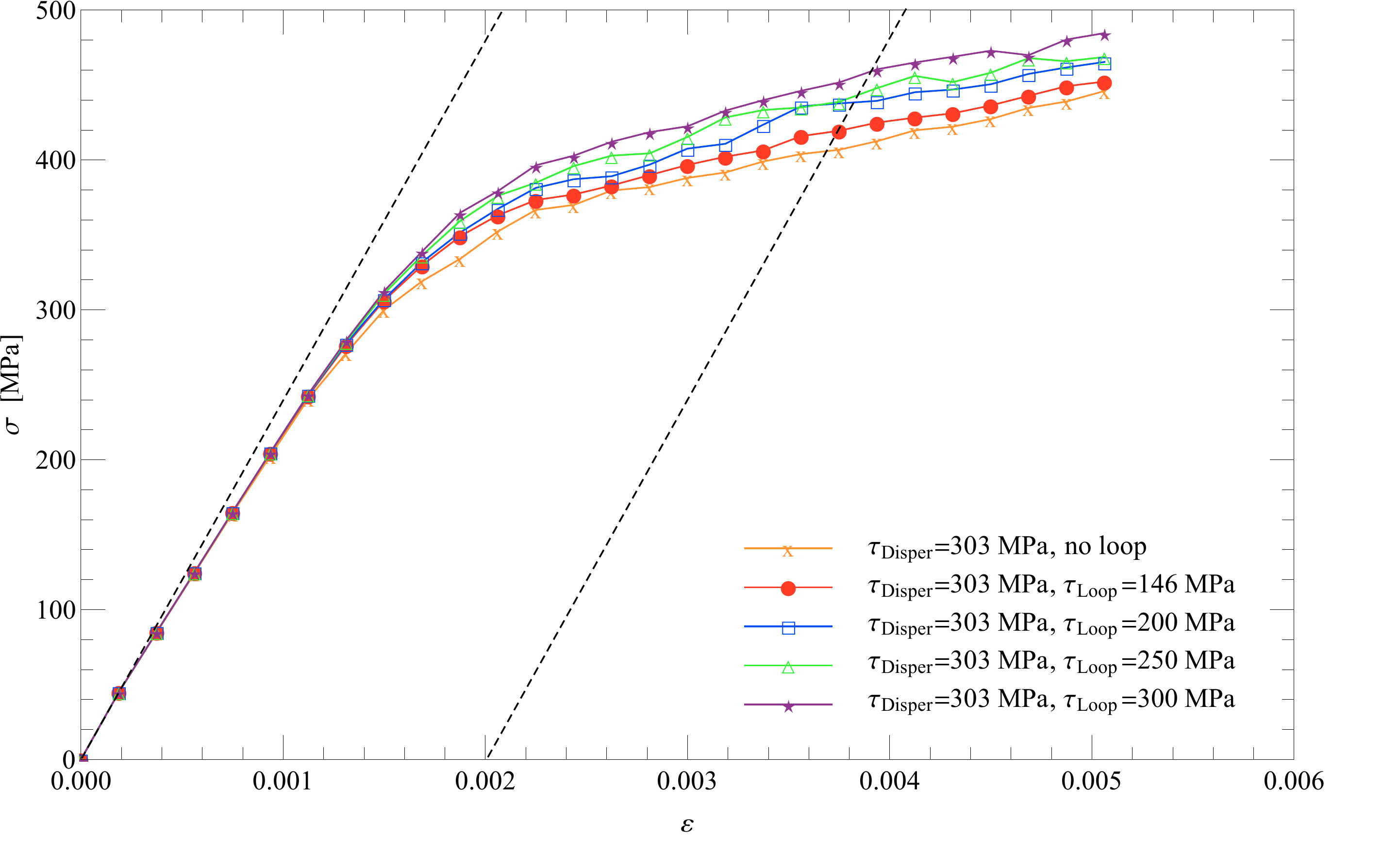}
\caption{Mixed hardening in the presence of dispersions with $\rho_{\rm Disper}^{\rm line}=3.72$ and prismatic loops with $\rho_{\rm Loop}^{\rm line}=1.74$ when $\alpha=\{0.33, 045, 0.57, 0.68\}$.}
\label{Mixedstrong}
\end{center}
\end{figure}
In Fig.~\ref{Mixedstrong}, it is shown that secondary defects may contribute effectively when their strength is not far from the strength of primary defects.
\subsection{Discussion}
The present study focuses on the extension of DRP which is based on the representation of defects in 2D DD by using fractional conservation laws and barrier hardening models. Existing 3D defects properties are translated to 1D basis with the aid of equivalence of fractions, and the resistance of each defect type is determined by using BKS or DBH model according to the observations based on the outcomes of external computational studies. However, any other suggested or modified strength expression may be preferred since these calculations are independent of 2D DD simulation framework. This setup is built on practical algorithms that provide realistic computational domains, control/track ability for its elements, and consequently opportunity to combine many properties by merely using points of interaction and analytical estimations.

Within the content of this study, it is seen that the geometrical properties of defects are extremely crucial in the governance of tensile characteristics since not only distribution of these objects, but also corresponding features depend on the size and density of defects. However, this information is supposed to be provided from either experimental or external computational studies, and DRP is liable to the existence and accuracy of defect properties. The present model has also limitations due to the nature of 2D DD. Screw dislocations have not been mentioned frequently because the dynamics of screw dislocations cannot be represented in 2D DD framework. Absence of mechanisms specific to screw dislocations such as kink-pair migration or cross-slip may cause inconsistency for BCC alloys for temperatures between 0-300 K since the plasticity of BCC materials is considered as governed by mainly screw dislocations. However, under the condition of plain strain deformation and in higher temperature intervals where screw dislocations are not mastering, barrier-dislocation interaction and related features may be suitable for simulations with the present framework.

Although the recipe of defect representation and principal features related to defects have been exhibited, enabling microstructural changes in 2D DD is started, but it is not over. Temperature dependence of the mobility laws is supposed to be implemented in either Arrhenius or non-Arrhenius form to efficiently test the performance of the materials although it requires a multi-scale approach for each material. There exists also substantial defect-concerned traits that can be easily done: grain boundary penetration by distributing finite strength obstacles on the interfaces of the grains, defect annihilation by dynamically setting the strength of corresponding defects to zero, or defect multiplication by initially distributing zero-strength defect points that will gain an effective strength in the further time steps of the simulation. 
\section{Conclusion}
The present study reveals out the following properties on the basis of the additional features of DRP for 2D DD:
\begin{itemize}
\item Representation of multi-type defects by using consecutively distributed points of interaction on slip lines shows consistency with the concentration-balanced superposition. Pre-determined barrier strength for each type of defect provides exceptional opportunity to speed up the simulations and to create flexibility for the usage of different barrier hardening formulations.
\item Dynamically increasing defect strength ends up with a significant rise in the yield strength. Pre-defined variation of barrier strength broadens the practicality of dimension reduction, and it leads to a strategy to deal with similar mechanisms such as defect annihilation. However, the parameters of these mechanisms may not be trivial and may require a careful treatment.
\item Static internal friction is an effective term which results in parallel shifts in the plastic response of the material. An enhanced function for friction may also be used to control the velocity of edge dislocations when necessary, but the origin of friction should be identified with lower scale methods or experimental data.  
\item A unique asset of DRP is the capability to perform large grain deformation. When the materials at the service of industry are taken into account, DRP for 2D DD is able to cope with realistic grain sizes as well as major heterogeneities in these grains. Therefore, DRP advances the feasibility of dislocation dynamics for a wide range of operative materials.  
\item The main disadvantages of present framework are the restriction to the plain strain deformation and the absence of screw dislocations. Plain strain deformation is the principal condition under which 2D DD is employed; however, deformation types which load screw dislocations are not suitable with 2D DD. Additionally, in the cases where yield strength strongly depends on the dominant mechanisms and properties related to screw dislocations, the outcome of DRP for 2D DD may be unsatisfying.
\item Modeling of other mechanical phenomena like Mode I crack propagation or 2D nanoindentation is also suitable for the application of DRP. 
\end{itemize}
\section*{Acknowledgments}
This research was carried out under project number M74.2.10411 in the framework of the Research Program of the Materials innovation institute M2i (www.m2i.nl). 
\bibliographystyle{ieeetr}
\nocite{}
\bibliography{2DDD,beff,Peierls,EUROFER,BCCAlloys,FCCAlloys,mixrule,fermar,austen,3DDDD}

\begin{thebibliography}{10}

\bibitem{Woolhouse1971}
G.~Woolhouse and M.~Ipohorski, ``On the interaction between radiation damage
  and coherent precipitates,'' {\em Proceedings of the Royal Society of London.
  Series A, Mathematical and Physical Sciences}, pp.~415--431, 1971.

\bibitem{Satoh1988}
Y.~Satoh, I.~Ishida, T.~Yoshiie, and M.~Kiritani, ``Defect structure
  development in 14 {MeV} neutron irradiated copper and copper dilute alloys,''
  {\em Journal of Nuclear Materials}, vol.~155, pp.~443--448, 1988.

\bibitem{Singh1996}
B.~N. Singh, D.~J. Edwards, and P.~Toft, ``Effects of neutron irradiation on
  mechanical properties and microstructures of dispersion and precipitation
  hardened copper alloys,'' {\em Journal of Nuclear Materials}, vol.~238,
  no.~2, pp.~244--259, 1996.

\bibitem{CuCrZr}
D.~J. Edwards, B.~N. Singh, and S.~T{\"a}htinen, ``Effect of heat treatments on
  precipitate microstructure and mechanical properties of a {CuCrZr} alloy,''
  {\em Journal of Nuclear Materials}, vol.~367, pp.~904--909, 2007.

\bibitem{Holmes1968}
J.~J. Holmes, R.~E. Robbins, J.~L. Brimhall, and B.~Mastel, ``Elevated
  temperature irradiation hardening in austenitic stainless steel,'' {\em Acta
  Metallurgica}, vol.~16, no.~7, pp.~955--967, 1968.

\bibitem{Lucas1993}
G.~E. Lucas, ``The evolution of mechanical property change in irradiated
  austenitic stainless steels,'' {\em Journal of Nuclear Materials}, vol.~206,
  no.~2, pp.~287--305, 1993.

\bibitem{Zinkle1993}
S.~J. Zinkle, P.~J. Maziasz, and R.~E. Stoller, ``Dose dependence of the
  microstructural evolution in neutron-irradiated austenitic stainless steel,''
  {\em Journal of Nuclear materials}, vol.~206, no.~2, pp.~266--286, 1993.

\bibitem{Hashimoto1999}
N.~Hashimoto, E.~Wakai, and J.~P. Robertson, ``Relationship between hardening
  and damage structure in austenitic stainless steel 316ln irradiated at low
  temperature in the hfir,'' {\em Journal of Nuclear Materials}, vol.~273,
  no.~1, pp.~95--101, 1999.

\bibitem{Jiao2010}
Z.~Jiao and G.~S. Was, ``The role of irradiated microstructure in the localized
  deformation of austenitic stainless steels,'' {\em Journal of Nuclear
  Materials}, vol.~407, no.~1, pp.~34--43, 2010.

\bibitem{Schaeublin2002}
R.~Schaeublin, D.~Gelles, and M.~Victoria, ``Microstructure of irradiated
  ferritic/martensitic steels in relation to mechanical properties,'' {\em
  Journal of Nuclear Materials}, vol.~307, pp.~197--202, 2002.

\bibitem{Klueh2007}
R.~L. Klueh and A.~T. Nelson, ``Ferritic/martensitic steels for next-generation
  reactors,'' {\em Journal of Nuclear Materials}, vol.~371, no.~1, pp.~37--52,
  2007.

\bibitem{Tanigawa2009}
H.~Tanigawa, R.~L. Klueh, N.~Hashimoto, and M.~A. Sokolov, ``Hardening
  mechanisms of reduced activation ferritic/martensitic steels irradiated at
  300$^\circ$ {C},'' {\em Journal of Nuclear Materials}, vol.~386,
  pp.~231--235, 2009.

\bibitem{Tanigawa2011}
H.~Tanigawa, K.~Shiba, A.~M{\"o}slang, R.~E. Stoller, R.~Lindau, M.~A. Sokolov,
  G.~R. Odette, R.~J. Kurtz, and S.~Jitsukawa, ``Status and key issues of
  reduced activation ferritic/martensitic steels as the structural material for
  a {DEMO} blanket,'' {\em Journal of Nuclear Materials}, vol.~417, no.~1,
  pp.~9--15, 2011.

\bibitem{Cockeram2011}
B.~V. Cockeram, R.~W. Smith, N.~Hashimoto, and L.~L. Snead, ``The swelling,
  microstructure, and hardening of wrought lcac, tzm, and ods molybdenum
  following neutron irradiation,'' {\em Journal of Nuclear Materials},
  vol.~418, no.~1, pp.~121--136, 2011.

\bibitem{Monnet2006}
G.~Monnet, ``Investigation of precipitation hardening by dislocation dynamics
  simulations,'' {\em Philosophical Magazine}, vol.~86, no.~36, pp.~5927--5941,
  2006.

\bibitem{Bako2007}
B.~Bako, D.~Weygand, M.~Samaras, J.~Chen, M.~A. Pouchon, P.~Gumbsch, and
  W.~Hoffelner, ``Discrete dislocation dynamics simulations of dislocation
  interactions with $\rm y_{2}o _{3}$ particles in {PM2000} single crystals,''
  {\em Philosophical Magazine}, vol.~87, no.~24, pp.~3645--3656, 2007.

\bibitem{Queyreau2010}
S.~Queyreau, G.~Monnet, and B.~Devincre, ``Orowan strengthening and forest
  hardening superposition examined by dislocation dynamics simulations,'' {\em
  Acta Materialia}, vol.~58, no.~17, pp.~5586--5595, 2010.

\bibitem{Robertson2011}
C.~Robertson and K.~Gururaj, ``Plastic deformation of ferritic grains in
  presence of {ODS} particles and irradiation-induced defect clusters:{ A 3D}
  dislocation dynamics simulation study,'' {\em Journal of Nuclear Materials},
  vol.~415, no.~2, pp.~167--178, 2011.

\bibitem{Topuz}
A.~I. Topuz, ``Dimension reduction of defect properties for application in {2D}
  dislocation dynamics,'' {\em Computational Materials Science}, vol.~95,
  pp.~13--20, 2014.

\bibitem{main}
E.~van~der Giessen and A.~Needleman, ``Discrete dislocation plasticity: a
  simple planar model,'' {\em Modelling Simui. Mater. Sci. Eng.}, vol.~3,
  pp.~689--735, 1995.

\bibitem{Ricebcc}
J.~R. Rice, ``Tensile crack tip fields in elastic-ideally plastic crystals,''
  {\em Mechanics of Materials}, vol.~6, no.~4, pp.~317--335, 1987.

\bibitem{Wang2009}
Y.~Wang, J.~W. Kysar, S.~Vukelic, and Y.~L. Yao, ``Spatially resolved
  characterization of geometrically necessary dislocation dependent deformation
  in microscale laser shock peening,'' {\em Journal of Manufacturing Science
  and Engineering}, vol.~131, no.~4, 2009.

\bibitem{Khan1999}
A.~S. Khan and R.~Liang, ``Behaviors of three bcc metal over a wide range of
  strain rates and temperatures: experiments and modeling,'' {\em International
  Journal of Plasticity}, vol.~15, no.~10, pp.~1089--1109, 1999.

\bibitem{Ramar2007}
A.~Ramar, N.~Baluc, and R.~Sch{\"a}ublin, ``Effect of irradiation on the
  microstructure and the mechanical properties of oxide dispersion strengthened
  low activation ferritic/martensitic steel,'' {\em Journal of Nuclear
  Materials}, vol.~367, pp.~217--221, 2007.

\bibitem{Heintze2011}
C.~Heintze, F.~Bergner, A.~Ulbricht, M.~Hern{\'a}ndez-Mayoral, U.~Keiderling,
  R.~Lindau, and T.~Weissg{\"a}rber, ``Microstructure of oxide dispersion
  strengthened eurofer and iron--chromium alloys investigated by means of
  small-angle neutron scattering and transmission electron microscopy,'' {\em
  Journal of Nuclear Materials}, vol.~416, no.~1, pp.~35--39, 2011.

\bibitem{Dele}
M.~A. Delesse, ``Proc\'{e}d\'{e}e m\'{e}canique pour d\'{e}terminer la
  composition des roches,'' {\em C. R. Acad. Sci. (Paris)}, vol.~25,
  pp.~544--545, 1847.

\bibitem{Rosi}
A.~Rosiwal, ``{\"U}ber geometrische gesteinsanalysen,'' {\em Verh. KK Geol.
  Reichsanst. Wien}, vol.~143, 1898.

\bibitem{DBH}
A.~K. Seeger, ``{2nd UN Conference on Peaceful Uses of Atomic Energy},'' {\em
  United Nations, New York}, vol.~6, p.~250, 1958.

\bibitem{BKS}
D.~J. Bacon, U.~F. Kocks, and R.~O. Scattergood, ``The effect of dislocation
  self-interaction on the {Orowan} stress,'' {\em Philosophical Magazine},
  vol.~28, no.~6, pp.~1241--1263, 1973.

\bibitem{Brown1971}
L.~M. Brown, R.~K. Ham, A.~Kelly, and R.~B. Nicholson, ``Strengthening methods
  in crystals,'' {\em Applied Science, London}, p.~9, 1971.

\bibitem{Lagerpusch2000}
U.~Lagerpusch, V.~Mohles, D.~Baither, B.~Anczykowski, and E.~Nembach, ``Double
  strengthening of copper by dissolved gold-atoms and by incoherent
  {SiO}$_{2}$-particles: how do the two strengthening contributions
  superimpose?,'' {\em Acta Materialia}, vol.~48, no.~14, pp.~3647--3656, 2000.

\bibitem{Hiratani2004}
M.~Hiratani and V.~V. Bulatov, ``Solid-solution hardening by point-like
  obstacles of different kinds,'' {\em Philosophical Magazine Letters},
  vol.~84, no.~7, pp.~461--470, 2004.

\bibitem{Kocks1980}
U.~F. Kocks, ``{Proceedings of the Fifth International Conference in the
  Strength of Metals and alloys, {27-31} August 1979 (Oxford, Aachen, Germany:
  Pergamon)},'' p.~1661, 1980.

\bibitem{Kang2012}
K.~Kang, V.~V. Bulatov, and W.~Cai, ``Singular orientations and faceted motion
  of dislocations in body-centered cubic crystals,'' {\em Proceedings of the
  National Academy of Sciences}, vol.~109, no.~38, pp.~15174--15178, 2012.

\bibitem{Kubin1998}
L.~P. Kubin, B.~Devincre, and M.~Tang, ``Mesoscopic modelling and simulation of
  plasticity in fcc and bcc crystals: Dislocation intersections and mobility,''
  {\em Journal of Computer-Aided Materials Design}, vol.~5, no.~1, pp.~31--54,
  1998.

\bibitem{Marian2004}
J.~Marian, W.~Cai, and V.~V. Bulatov, ``Dynamic transitions from smooth to
  rough to twinning in dislocation motion,'' {\em Nature Materials}, vol.~3,
  no.~3, pp.~158--163, 2004.

\bibitem{Edwards2005}
D.~J. Edwards, B.~N. Singh, and J.~Bilde-S{\o}rensen, ``Initiation and
  propagation of cleared channels in neutron-irradiated pure copper and a
  precipitation hardened {CuCrZr} alloy,'' {\em Journal of Nuclear Materials},
  vol.~342, no.~1, pp.~164--178, 2005.

\bibitem{Byun2008}
T.~S. Byun and S.~A. Maloy, ``Dose dependence of mechanical properties in
  tantalum and tantalum alloys after low temperature irradiation,'' {\em
  Journal of Nuclear Materials}, vol.~377, no.~1, pp.~72--79, 2008.

\bibitem{Wang2011}
Z.~Q. Wang and I.~J. Beyerlein, ``An atomistically-informed dislocation
  dynamics model for the plastic anisotropy and tension--compression asymmetry
  of bcc metals,'' {\em International Journal of Plasticity}, vol.~27, no.~10,
  pp.~1471--1484, 2011.

\bibitem{Schaublin2006}
R.~Sch{\"a}ublin, A.~Ramar, N.~Baluc, V.~De~Castro, M.~Monge, T.~Leguey,
  N.~Schmid, and C.~Bonjour, ``Microstructural development under irradiation in
  {European} {ODS} ferritic/martensitic steels,'' {\em Journal of Nuclear
  Materials}, vol.~351, no.~1, pp.~247--260, 2006.

\bibitem{Lindau2002}
R.~Lindau, A.~M{\"o}slang, M.~Schirra, P.~Schlossmacher, and M.~Klimenkov,
  ``Mechanical and microstructural properties of a hipped {RAFM ODS-steel},''
  {\em Journal of Nuclear Materials}, vol.~307, pp.~769--772, 2002.

\bibitem{Eiselt2009}
C.~C. Eiselt, M.~Klimenkov, R.~Lindau, A.~M{\"o}slang, H.~Sandim, A.~Padilha,
  and D.~Raabe, ``High-resolution transmission electron microscopy and electron
  backscatter diffraction in nanoscaled ferritic and ferritic--martensitic
  oxide dispersion strengthened--steels,'' {\em Journal of Nuclear Materials},
  vol.~385, no.~2, pp.~231--235, 2009.

\bibitem{Odette1979}
G.~R. Odette and D.~Frey, ``Development of mechanical property correlation
  methodology for fusion environments,'' {\em Journal of Nuclear Materials},
  vol.~85, pp.~817--822, 1979.

\end{thebibliography}
\end{document}